\documentclass[fleqn,usenatbib]{mnras}
\usepackage[latin1]{inputenc}
\usepackage[english]{babel}
\usepackage[T1]{fontenc}
\usepackage{ae,aecompl}
\usepackage{graphicx}
\usepackage{xspace}
\usepackage{amsmath}
\usepackage{amssymb}
\usepackage{epsfig}
\usepackage{pdflscape}
\usepackage{subcaption}
\usepackage{verbatim}
\usepackage{amsfonts}
\usepackage{hyperref}
\usepackage{color}
\usepackage{float}
\usepackage{url}
\usepackage{booktabs}
\usepackage{fancyhdr}
\usepackage{multirow}
\usepackage{grffile}
\usepackage{geometry}
\newcommand{\cmb}{CMB }
\newcommand{\pl}{Planck}
\newcommand{\mcmole}{\texttt{MCMole3D}}
\newcommand{\axis}{\texttt{Axisymmetric}}
\newcommand{\spir}{\texttt{LogSpiral}}
\newcommand{\xpure}{\texttt{X$^{2}$PURE}}

\title[A 3D model for CO  emission]{A 3D model for CO  molecular line emission  as a potential \cmb polarization contaminant}

\author[G. Puglisi et al.]{
G. Puglisi,$^{1,2}$\thanks{E-mail: \url{giuspugl@sissa.it}}
G. Fabbian,$^{3,1,2}$\thanks{E-mail: \url{giulio.fabbian@ias.u-psud.fr}}
C. Baccigalupi$^{1,2}$
\\
$^{1}$SISSA- International School for Advanced Studies, Via Bonomea 265, 34136 Trieste, Italy\\
$^2$INFN-National Institute for Nuclear Physics, Via Valerio 2, I-34127 Trieste, Italy\\
$^{3}$Institut d'Astrophysique Spatiale, CNRS (UMR 8617), Univ. Paris-Sud, Universit\'{e} Paris-Saclay, b\^{a}t. 121, 91405 Orsay, France\\
}

\date{Accepted XXX. Received YYY; in original form ZZZ}
\pubyear{2016}

\begin{document}

\label{firstpage}
\pagerange{\pageref{firstpage}--\pageref{lastpage}}
\maketitle
\begin{abstract}
 We present a model for simulating Carbon Monoxide (CO) rotational line 
emission in molecular clouds, taking account of their 3D spatial 
distribution in galaxies with different geometrical properties. The model 
implemented is based on recent results in the literature and has been 
designed for performing Monte-Carlo simulations of this emission. We 
compare the simulations produced with this model and calibrate them, both 
on the map level and on the power spectrum level, using the second release 
of data from the \pl\ satellite for the Galactic plane, where the 
signal-to-noise ratio is highest. We use the calibrated model to 
extrapolate the CO power spectrum at low Galactic latitudes where no high 
sensitivity observations are available yet. We then forecast the level of 
unresolved polarized emission from CO molecular clouds which could 
contaminate the power spectrum of Cosmic Microwave Background (CMB) 
polarization B-modes away from the Galactic plane. Assuming realistic 
levels of the polarization fraction, we show that the level of 
contamination is equivalent to a cosmological signal with $r \lesssim 
0.02$. The Monte-Carlo MOlecular Line Emission (\mcmole) \texttt{Python} 
package, which implements this model, is being made publicly available.
\end{abstract}

\begin{keywords}
\small
Interstellar Medium: molecules, magnetic fields, lines and bands  
Cosmology: observations, cosmic background radiation 
\end{keywords}

\section{Introduction}
\label{sec:intro}
The Carbon monoxide (CO) molecule is one of the most interesting molecules
present in molecular clouds within our Galaxy.  Although the most abundant
molecule in Galactic molecular clouds is molecular hydrogen ($\mathrm{H_2}$),
it is inconvenient to use the emission from that as a tracer since it is
difficult to detect because of having a low dipole moment and so being a very
inefficient radiator. We therefore need  to resort to alternative techniques
for tracing molecular clouds using rotational or vibrational transitions of
other molecules such as CO.  Observations of CO emission are commonly used
to infer the mass of molecular gas in the Milky Way by assuming a linear
proportionality between the CO and $H_2$ densities via the CO-to-$\mathrm{H}_2$
conversion factor, $X_{CO}$. A commonly accepted value for $X_{CO}$ is $
2\times 10^{20}\, \mathrm{molecules\cdot cm^{-2}\, \left(K\, km
\,s^{-1}\right)^{-1} }$, although this could vary with position in  the
Galactic plane, particularly in the outer Galaxy \citep{2011ApJ...738...27B}.
\\*

The most intense  CO rotational transition lines are the $J = 1\rightarrow0,\,
2\rightarrow1,\, 3\rightarrow2$ transitions at sub-millimetre wavelengths (115,
230 and 345 GHz respectively).  These can  usually be observed in optically
thick and thermalized regions of the interstellar medium.  Traditionally, the
observations of standard $\mathrm{{}^{12}CO}$ emission are complemented by
measurements of  $\mathrm{{}^{13}CO}$ lines. Being less abundant (few percent),
this isotopologue  can be exploited for inferring  the dust extinction in
nearby clouds and hence providing a better constraint for measuring the 
$\mathrm{H_2}$ abundance \citep{1987ApJ...312L..45B, 2006ApJS..163..145J}.
However, there is growing evidence that  $\mathrm{{}^{13}CO}$ regions could be
associated with  colder and denser environments, whereas $\mathrm{{}^{12}CO}$
emission originates from a diffuse component of molecular gas
\citep{Roman-Duval2016}.

\begin{figure} 
\includegraphics[width=\columnwidth, angle=180,trim= 0 1.2cm 0 0cm,clip=true]{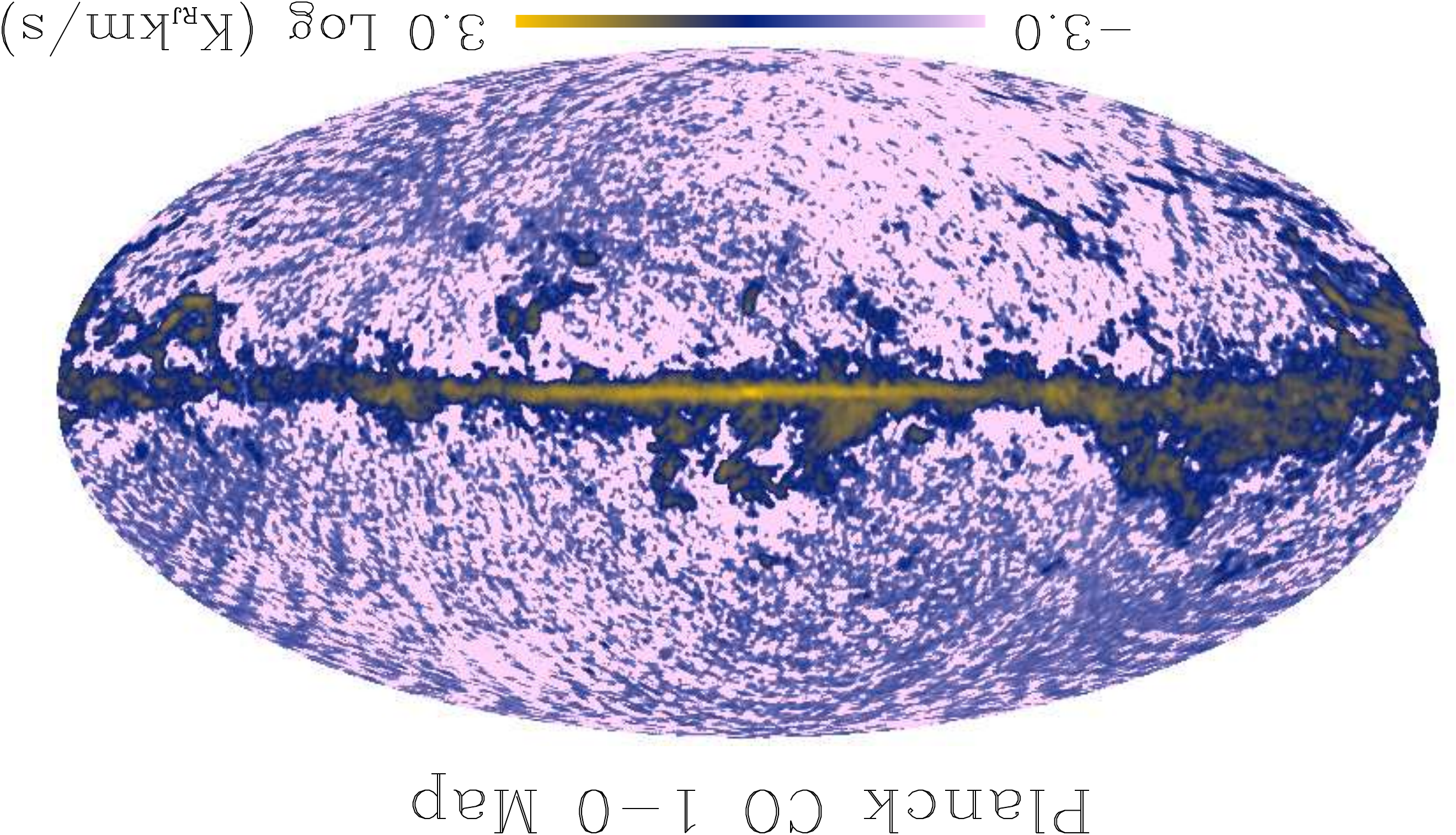}
\caption{ Planck CO $1-0$ map  \citep{plX2015}. Note the predominance of instrumental noise in regions far from the Galactic plane.
} \label{fig:cosurveys}
\end{figure}

 The spatial distribution of the CO line emission reaches a peak in the inner
Galaxy and is mostly concentrated close to or within the spiral arms, in a
well-defined ring, the so-called \emph{molecular ring} between about $4 - 7$
kpc from the Galactic centre. This property is not unique to the Milky Way but
is quite common in barred spiral galaxies (see \cite{2002ApJ...574..126R}  for
further references). The emission in the direction orthogonal to the Galactic
plane is confined within a Gaussian slab with roughly 90 pc full width half
maximum (FWHM) in the inner Galaxy getting broader towards the outer Galactic
regions, reaching a FWHM of several hundred parsecs outside the solar
circle. 
In the centre of the Galaxy, we can also identify a very dense CO emission zone
,rich in neutral gas and individual stars, stretching out to about 700 light
years (ly) from the centre and known as the \emph{Central Molecular Zone}.\\*
Since the  1970s, many CO surveys of the Galactic plane have been carried out
with ground-based telescopes, leading to accurate catalogues of molecular
clouds \citep{2001ApJ...547..792D, 2004ASPC..317...59M}. Usually these surveys
have observed a strip of $|b|\lesssim 5 \deg $ around the Galactic plane. At
higher Galactic latitudes ($|b|>30 \deg$), the low opacity regions of both gas
and dust, together with a relatively low stellar background which is useful for
spotting extinction regions, complicate the observation of CO lines making this
very challenging. In fact, only  $\approx100$ clouds have been detected so far
in these regions.\\*
The Planck satellite team recently released CO emission maps of the lowest
rotational  lines, $J = 1-0, \,2-1\,,3-2$ observed in the 100, 217, 353 GHz
frequency channels of the High Frequency Instrument (HFI)
\citep{2014A&A...571A..13P, plX2015}. These were sensitive enough to map the CO
emission even though the widths of these lines are orders of magnitude narrower
than the bandwidth of the \pl\ frequency channels. These single frequency maps
have been processed with a dedicated foreground cleaning procedure so as to
isolate this emission. The \pl\ maps were found to be broadly consistent with
the data from other CO surveys \citep{2001ApJ...547..792D, Heyer2015}, although
they might still be affected by residual astrophysical emissions and
instrumental systematics.  In \autoref{fig:cosurveys}, we show the so called
Type 1 \pl\ map of the CO $J:1-0$ line  \citet{2014A&A...571A..13P}\footnote{http://pla.esac.esa.int/pla} which will be
used in the following.\\

Many current and future CMB polarization experiments\footnote{For a complete
list of the operating and planned probes see e.g. \url{lambda.gfsc.nasa.gov}}
are designed to exploit the faint B-mode signal of CMB polarization as a
cosmological probe, in particular to constrain the physics of large scale
structure formation or the inflationary mechanism in the early universe
\citep{1997PhRvL..78.2054S,Hu:1997hv}.  One of the main challenges in the way
of achieving these goals is the contamination of the primordial CMB signal by
diffuse Galactic emission. In this respect, the synchrotron and thermal dust
emission are known to be potentially the most dangerous contaminants, because
they are intrinsically polarized. In fact, several analyses conducted on Planck
and Wilkinson Microwave Anisotropy Probe (WMAP) data from intermediate and high
Galactic latitudes at high \citep{PlanckCollaboration2014a} and low frequencies
\citep{Krachmalnicoff2015, 2016A&A...594A..25P} showed that these emissions are
dangerous at all microwave frequencies and locations on the sky (even if far
from the galactic plane), confirming early studies using the  WMAP satellite
\citep{2011ApJS..192...15G,2007ApJS..170..335P,2003NewAR..47.1127B}. \\*
Appropriate observations and theoretical investigations and modelling of polarized foreground emission for all emissions at sub-mm frequencies are
therefore crucial for the success of future experiments. As these will observe
at frequencies overlapping with the CO lines, unresolved CO line emission could
significantly contaminate these measurements as well. \\*
CO lines are in fact expected to be polarized at the percent level or below
\citep{Goldreich1981} because of interaction of the  magnetic moment of the molecule
with the Galactic magnetic field. This causes the so-called \emph{Zeeman
splitting} of the rotational quantum levels $J$ into the magnetic sub-levels
$M$ which are intrinsically polarized. Moreover, if molecular clouds are
somehow anisotropic (e.g when in the presence of expanding or collapsing envelopes
in star formation regions) or are asymmetric, population imbalances of the $M$
levels can arise. This leads to different line intensities depending on the
directions (parallel or perpendicular to the magnetic field) and to a net
linearly polarized emission.  \citet{Greaves1999} detected polarization in five
 star-forming regions near to the Galactic Centre while observing
the CO lines $J=2-1,\, 3-2$ and the $J=2-1$ of the isotopologue
${}^{13}\mathrm{CO} $. The degree of polarization ranged from $0.5
\mathrm{\,to\,} 2.5 \%$. Moreover, the deduced magnetic field direction was
found to be consistent with previous measurements coming from dust polarimetry,
showing that the polarized CO emission could become a sensitive tracer of 
small-scale Galactic magnetic fields. \\

The goal of this paper is to propose a statistical 3D parametric model of CO
molecular cloud emission, in order to forecast the contamination of CMB signal
by this, including in the polarization. Being able to perform statistical
simulation of this emission is crucial for assessing the impact of foreground residual uncertainties on cosmological constraints coming from the CMB.  In addition, the capability of modeling the Galactic foreground emission in its full complexity taking into account line-of-sight effects is becoming necessary in light of the latest experimental results and the expected level of sensitivity for the future experiments \citep{tassis2015, planck2017-L}. In \autoref{sec:mod} we present the assumptions made for building the model and the simulation pipeline for its implementation. In
\autoref{sec:compars} we describe the methodology for calibrating the CO
simulations to match \pl\ observations. 
In \autoref{sec:bestfit} we show how the parameters describing molecular cloud
distribution shape the angular power spectrum of CO emission. Finally, in
\autoref{sec:forecasts} we forecast the expected level of polarized CO
contaminations for  the \cmb\ B-modes at high Galactic latitudes using our
calibrated simulation of \autoref{sec:compars} to infer statistically the
emission at high Galactic latitude, where current observations are less
reliable.

\section{Building a statistical 3D CO emission model}\label{sec:mod}

In order to build an accurate description of CO emission in the Galaxy, we
collected the most up to date astrophysical data present in the literature
concerning the distribution of molecular gas as a function of the Galactic
radius ($R$) and the vertical scale of the Galactic disk ($z$) as well as of
the molecular size and the mass function.  The model has been implemented in a
Python package named \mcmole\footnote{https://github.com/giuspugl/MCMole3D}
which is being made publicly available, and we present  details  of it in this
Section\footnote{In the following we will refer to this model as the \mcmole\
model for the  sake of clarity.}. The model builds on and extends the method
proposed by \citet{Ellsworth-Bowers2015} who conducted a series of analyses
distributing statistically a relative large number of molecular cloud objects
according to the axisymmetric distribution of $\mathrm{H}_2$ observed in the
Galaxy \citep{2003ApJ...587..278W}.
\subsection{CO cloud spatial distribution}

\begin{figure*}
\begin{flushleft}
\subcaptionbox{}[.5\columnwidth]{\includegraphics[scale=.36,clip=true, trim=2cm 6cm 7cm 7cm]{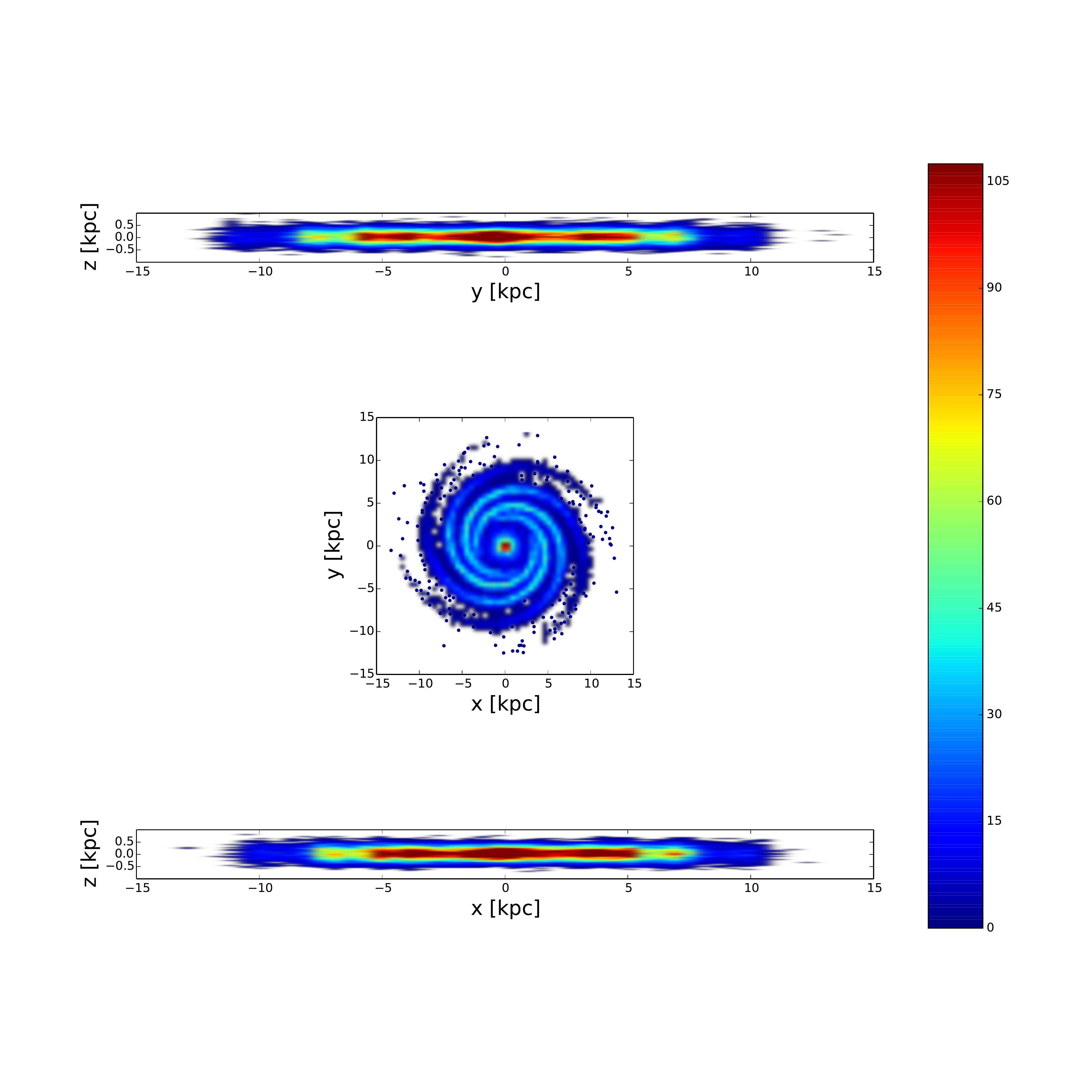}}\hspace{5cm}
\subcaptionbox{}[.5\columnwidth]{\includegraphics[scale=.36,clip=true, trim=4.7cm 6cm 7cm 7cm]{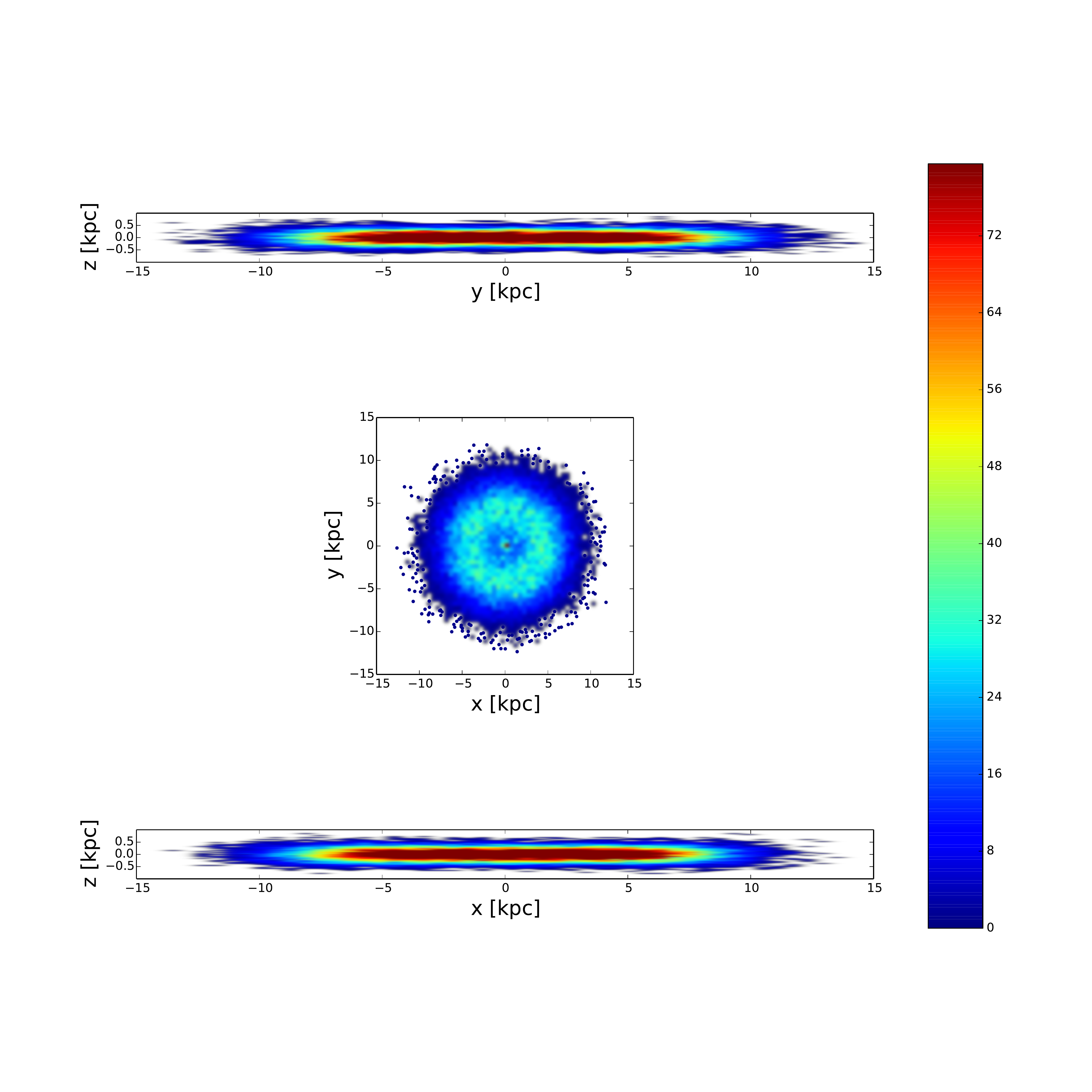}}\\
\subcaptionbox{}[.5\columnwidth]{\includegraphics[scale=.35,clip=true, trim=2cm 1cm 1cm 2cm]{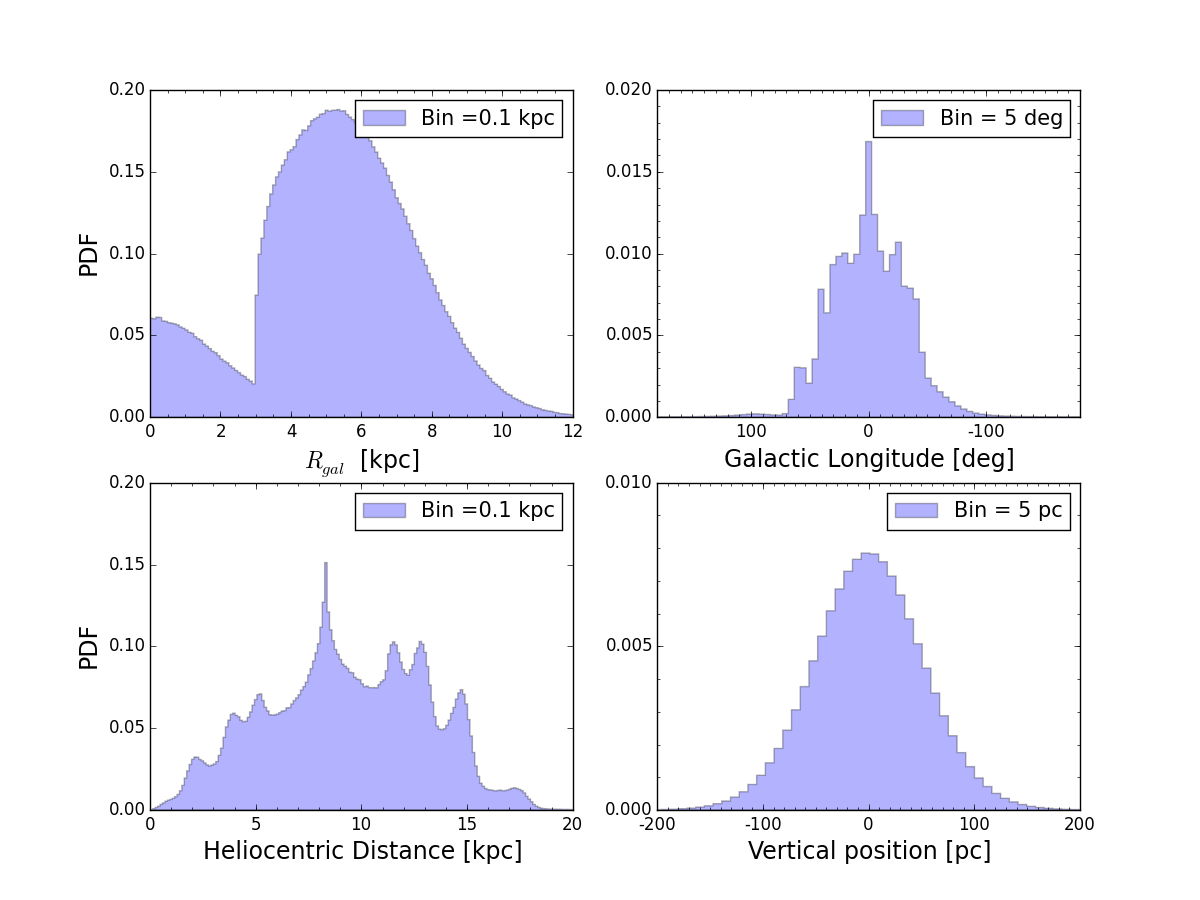}}\hspace{5.5cm}
\subcaptionbox{}[.5\columnwidth]{\includegraphics[scale=.35,clip=true, trim=2.5cm 1cm 1cm 1cm]{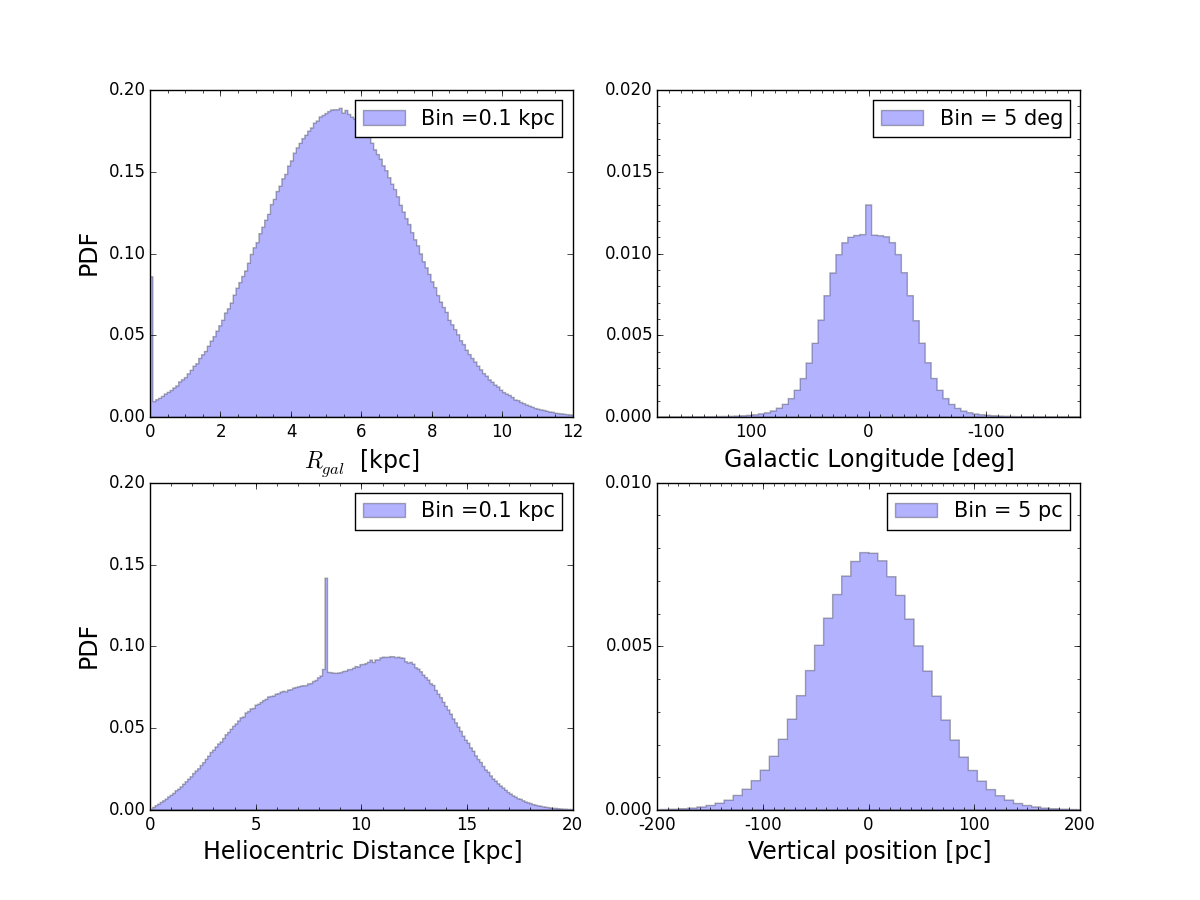}}
\caption{\emph{Top panels}: Density contour plots of an MC galaxy population with 40,000 objects distributed following the (a) \spir\ and (b) \axis\ distributions. \emph{Bottom panels}: Probability Density Function (PDF) of 100 MC realizations of 40,000 molecular clouds following the (c) \spir\ and (d) \axis\ geometry. The latter case is consistent with results in \citet{Ellsworth-Bowers2015}.}\label{fig:dens}
\end{flushleft}
\end{figure*}

As mentioned in the introduction, the CO emission is mostly concentrated around the molecular ring. We  have considered and implemented two different spatial distributions of the molecular clouds: an axisymmetric \emph{ring-shaped} one and one with 4 spiral arms, as shown in \autoref{fig:dens}(b) and (a) respectively.  
The first is a simplified model and is parametrized by $R_{ring}$,and $ \sigma_{ring}$ which are  the radius and the width of the molecular ring respectively. 
On the other hand, the spiral arm distribution is in principle closer to the symmetry of our Galaxy and is therefore more directly related to observations. The distribution is described by two more parameters than for the axisymmetric case: the arm width and the spiral arm pitch angle. For the analysis conducted in the following sections, we fixed the value of the pitch angle to be $i \sim -13\deg$  following the latest measurements of \citet{Davis2012, Bobylev2013} and fixed the arm half-width to be $340 $ pc \citep{Vallee2014}.\\*
\citet{Bronfman1988} found that the vertical profile of the  CO emissivity  can be optimally described by a Gaussian function of $z$ centred on $z_0$  and having a half-width $ z_{1/2}$ from the Galactic plane at $z=0$. Both of the parameters $z_0$ and $z_{1/2}$ are in general functions of the Galactic radius $R$ (see \citet{Heyer2015} for recent measurements).  
Since we are interested in the overall distribution of molecular clouds mainly in regions close to the Galactic plane, where data are more reliable, we adopted this parametrization but neglected the effects of the mid-plane displacement $z_0$ and set it to a constant value $z_0=0$, following \citet{2013A&A...553A..96D}. The vertical profile is then parametrized just by $z_{1/2} $ and mimics the increase of the vertical thickness scatter that is observed when moving from the inner Galaxy towards the outer regions: 
\begin{equation}
z_{1/2}(R)\propto \sigma_z(R)=\sigma_{z,0} \cosh \left(\frac{R}{h_R } \right),
\end{equation}
where $\sigma_{z,0} =0.1 $ and $h_R = 9$ kpc corresponds to the radius where the vertical thickness starts increasing. The half-width $z_{1/2} $ is related to $\sigma_z$ through the usual relation $ z_{1/2}=\sqrt{2 \ln 2}\sigma_z$.
The final vertical profile is then:
 \begin{equation}
 z(R) = \frac{1}{\sqrt{2 \pi} \sigma_z(R) } \exp \left[- \left(\frac{z}{\sqrt{2} \sigma_z(R) }\right)^2 \right] . \label{eq:zfunc}
\end{equation}

\begin{figure}
\includegraphics[width=\columnwidth]{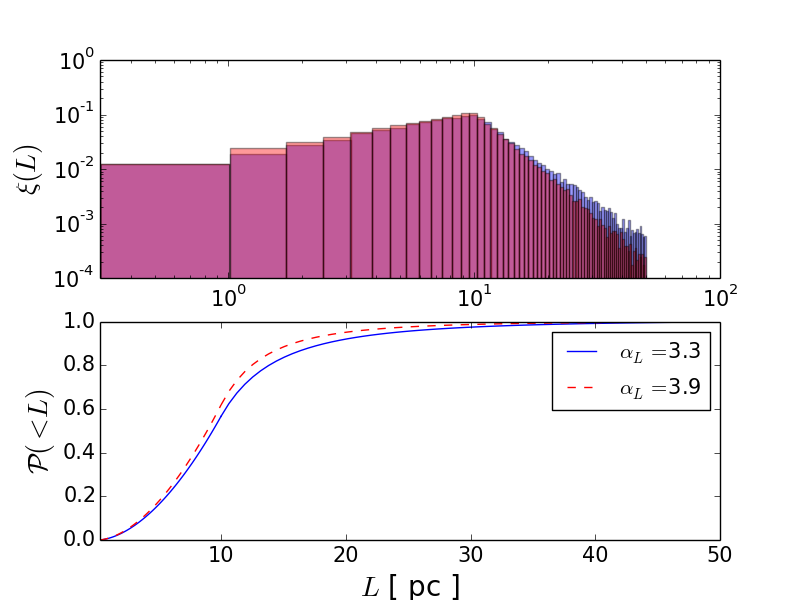}
\caption{(top) Histograms of $dN/dL$ computed by assigning the size of  each cloud with the probability function (bottom). The two spectral indices $\alpha_L\approx 3.3 \left(3.9\right) $ refer respectively to clouds in the inner (outer) Galaxy.}\label{fig:sizefunc}
\end{figure}

\subsection{CO cloud emission}
The key ingredients for modeling  the molecular cloud emission are the dimension of the cloud and its typical emissivity. We assume an exponential CO emissivity profile which  is a function of the Galactic radius following \citet{Heyer2015,Roman-Duval2016}:  
\begin{equation}
\epsilon_0(R)=\epsilon_c \exp \left(R/R_{em} \right), \label{eq:emprof}
\end{equation}
where $\epsilon_c$ is the typical emissivity of a particular CO line observed towards the centre of the Galaxy and $R_{em}$ the scale length over which the emissivity profile changes. Clouds observed in the outer Galaxy are in fact dimmer.\\*
We then assume the distribution of cloud size $\xi(L)$ defined by their typical size scale, $L_0 $, the range of sizes $\left[L_{min}, L_{max}\right]$ and two power-laws with spectral indices \citep{Roman-Duval2010}
\begin{equation}
\xi(L)=\frac{dn}{dL}\propto  
\left\{
	\begin{array}{ll}
		L^{0.8}  & \mbox{if } L_{min}<L< L_0,\\
		L^{-\alpha_L}  & \mbox{if } L_0<L< L_{max},
	\end{array}
\right.
\end{equation}
with $\alpha_L= 3.3, 3.9 $ for clouds inside or outside the solar circle respectively. From the cloud size function $\xi(L)$ we derive the corresponding probability $\mathcal{P}(L)$ of having clouds with sizes smaller than $L$:
\begin{equation}
\mathcal{P}(<L)=\int_{L_{min}} ^{L} d {L'} \xi(L') .
\label{eq: probfunc}
\end{equation}  
The probability functions for different choices of the spectral index $\alpha_L$ are shown in \autoref{fig:sizefunc}. 
We then inverted \autoref{eq: probfunc} to get the cloud size associated with a given probability $L(p)$. The cloud sizes are drawn from a uniform distribution in $\left[ 0,1\right]$. The histograms of the sizes generated following this probability function are shown in the top panel of \autoref{fig:sizefunc} and are peaked around the most typical size  $L_0$. In the analysis presented in the following $L_0$ is considered as a free parameter.\\*
Finally, we assume a spherical shape for each of the  simulated molecular clouds once they are projected on the sky. However, we implemented different emissivity profiles that are function of the distance from the cloud center, such as \emph{Gaussian} or \emph{cosine} profiles. These are particularly useful because, by construction, they give zero emissivity at the  boundaries\footnote{For the Gaussian profile, we set $\sigma$ in order to have the cloud boundaries at $6\sigma$, i.e. where the Gaussian function is zero to numerical precision.} and the maximum of the emissivity in the centre of the projected cloud on the sky. This not only mimics a decrease of the emission towards the outer regions of the cloud, where the density decreases, but also allows to minimize numerical artifacts when computing the angular power spectrum of the simulated maps (see \autoref{sec:compars}). An abrupt top-hat transition at the boundary of each cloud would in fact cause ringing effects that could bias the estimate of the power spectrum. 

\begin{table}
\footnotesize
\begin{tabular}{ll}
\toprule
\mcmole\ Default parameters\\
\midrule
$N_{clouds}$  & 40,000\\
$R_{ring}$ [kpc]& 5.3\\
$L_{min}$ [pc] &0.3\\
$L_{max}$ [pc] &60 \\ 
$\sigma_{z,0}$ [pc] & 100 \\
$h_R$ [kpc] &  9 \\
$R_{bar} \,^{\dag}$ [kpc] & 3 \\
$i \,^{\dag}$ [deg] & -12 \\
$\epsilon_c\, \left[\mathrm{K_{RJ} \,km \,s^{-1}} \right]$ &240 \\
$R_{em}$ [kpc] & 6.6 \\
\midrule
$L_{0} $ [pc] & [5,50] Default: 20\\
$\sigma_{ring} $ [kpc] & [1,5] Default:2.5\\
\bottomrule
\end{tabular}
\caption{List of  parameters used in \mcmole\ simulations. $ {}^\dag$ only  for \spir.}\label{tab:params}
\end{table}

\subsection{Simulation procedure}\label{sec:simproc}
The model outlined in the previous Section enables statistical simulations of CO emission in our Galaxy to be performed for a given set of free parameters  $\Theta^{CO}$ that can be set by the user:
\begin{align*}
 \Theta^{CO}= \lbrace & N_{clouds},\, \epsilon_c,\,R_{em},\, R_{ring},\, \sigma_{ring},\\    
  & \sigma_{z,0},\,h_R,\,  L_{min},L_{max}, L_0\rbrace. 
\end{align*}
The values chosen for our analysis are listed in \autoref{tab:params}.
\begin{figure*}
\includegraphics[width=\columnwidth , angle=180,trim= 0 0cm 0 0cm,clip=true]{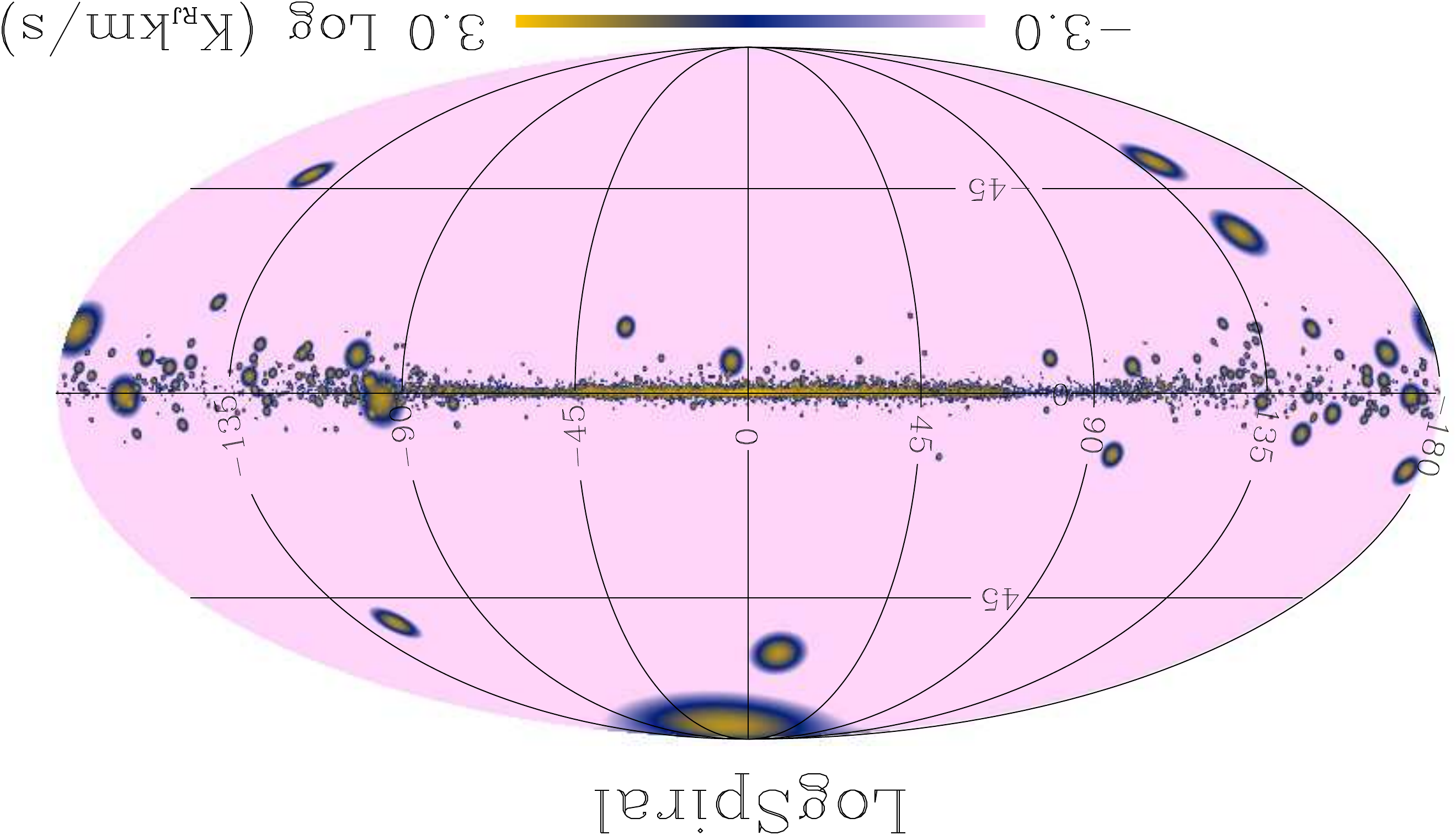}
\includegraphics[width=\columnwidth , angle=180,trim= 0 0cm 0 0cm,clip=true]{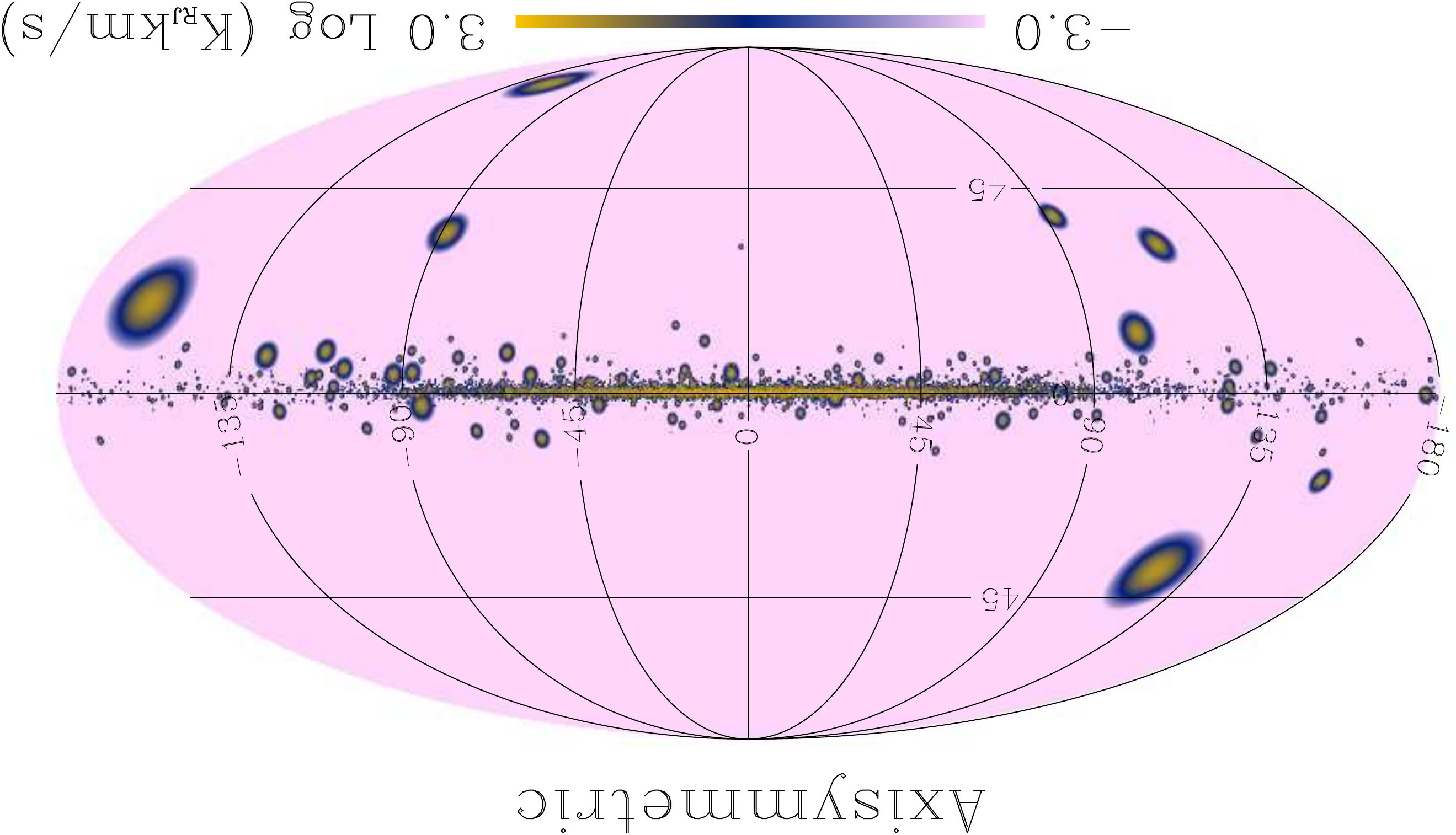}
\caption{Two realizations  of CO maps simulated with \mcmole\ using the distribution parameters given by the values in \autoref{tab:params}.} \label{fig:mapsim}
\end{figure*}
\noindent
For each realization of the model, we distribute by default 40,000 clouds within our Galaxy. This number is adopted for consistency with observations when observational cuts are applied (for further details see \citet{Ellsworth-Bowers2015}). The product of each simulation is a map, similar to the one in \autoref{fig:mapsim}, in the Hierarchical Equal Area Latitute Pixelization  (HEALP{\scriptsize IX}, \citet{2005ApJ...622..759G} ) \footnote{http://healpix.sourceforge.net} pixelization scheme including all the simulated clouds as seen by an observer placed in the solar system. This map can be smoothed to match the resolution of a specific experiment and/or convolved with a realistic frequency bandwidth. When we compare with the \pl\ maps described in \autoref{sec:compars}, we convolve the simulated maps to the beam resolution of the 100 GHz channel ($\sim 10 $ arcmin).\\* 
The procedure implemented for each realization is the following: 
\begin{enumerate}
\item assign the $(R_{gal},\phi,z)$ Galacto-centric positions. In particular:
\begin{itemize}
\item $R_{gal}$ is extracted from a Gaussian distribution defined by  the $R_{ring}$ and $\sigma_{ring}$  parameters. However, the $\sigma_{ring}$ is large enough to give non-zero  probability at $R_{gal}\leq 0$. All of the negative values of $R_{gal}$ are either automatically set to $R_{gal}=0$ (axisymmetric case), or  recomputed extracting new positive values from a normal distribution centred at $R=0$ and with the r.m.s given by the scale of the Galactic bar (spiral-arm case). This choice allows us to circumvent not only the issue of negative values of $R_{gal}$ due to a Gaussian distribution, but also to produce the high emissivity of the Central Molecular zone (see \citet{Ellsworth-Bowers2015}  for a similar approach). 

\item the $z$-coordinate is drawn randomly from the distribution in \autoref{eq:zfunc}.
 
\item the azimuth angle $\phi $ is computed from a uniform distribution ranging over $[0,2\pi )$ in the case of the axial symmetry. Conversely, in the case of spiral arms, $\phi$ follows the \emph{logarithmic spiral } polar equation 
\begin{displaymath}
\phi(R)= A \log R +B , 
\end{displaymath}
where $A=(\tan i)^{-1}$ and $B=-\log R_{bar}$ are, respectively, functions of the mean pitch angle and the starting radius of the spiral arm. In our case we set $i=-12$ deg, $R_{bar}=3$ kpc;
\end{itemize} 
\item assign cloud sizes given  the probability function $\mathcal{P}(L)$ (\autoref{eq: probfunc});
\item assign emissivities to each cloud from the emissivity profile (see \autoref{eq:emprof});
\item convert $(R_{gal},\phi,z)$ positions into the \emph{heliocentric } coordinate frame $(\ell,b,d_{\odot})$. 
\end{enumerate}
In \autoref{fig:dens} we show an example of the 3D distribution of the emission as well as the distribution of the location of the simulated clouds using both of the geometries implemented in the package.

\begin{figure*}
\subcaptionbox{}[\columnwidth]{\includegraphics[width=\columnwidth]{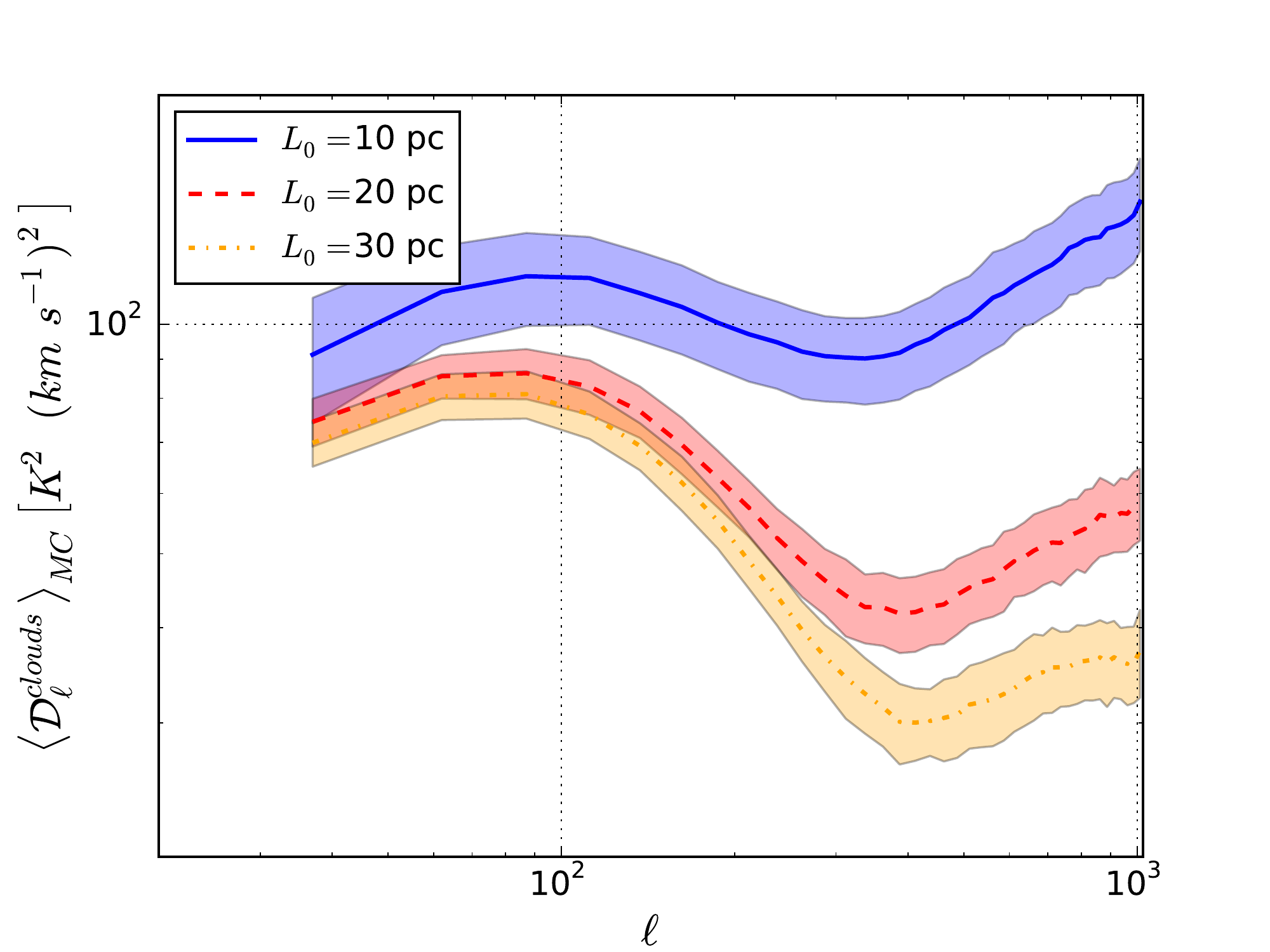}}
\subcaptionbox{}[\columnwidth]{\includegraphics[width=\columnwidth]{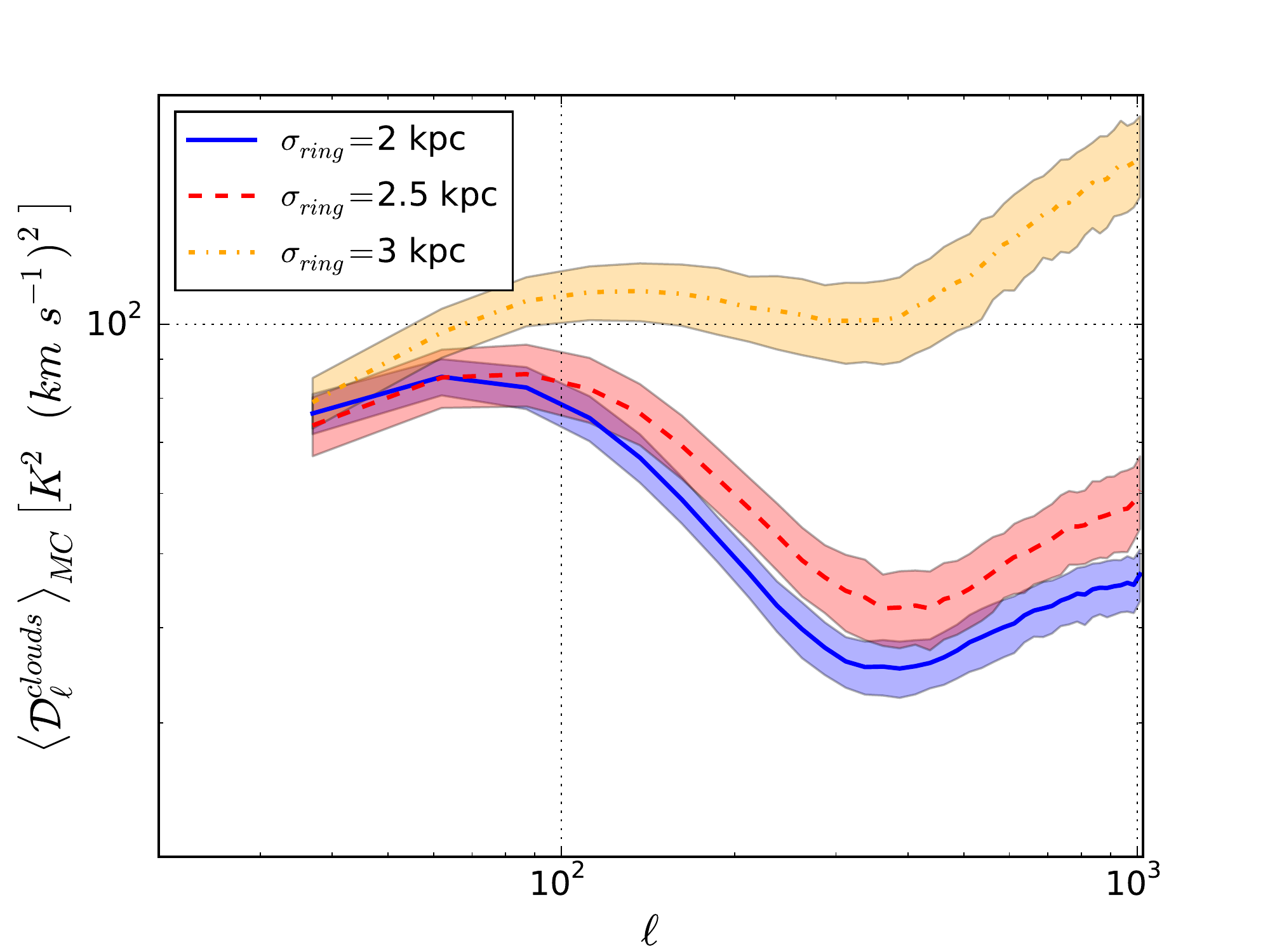}}\\
\subcaptionbox{}[\columnwidth]{\includegraphics[width=\columnwidth]{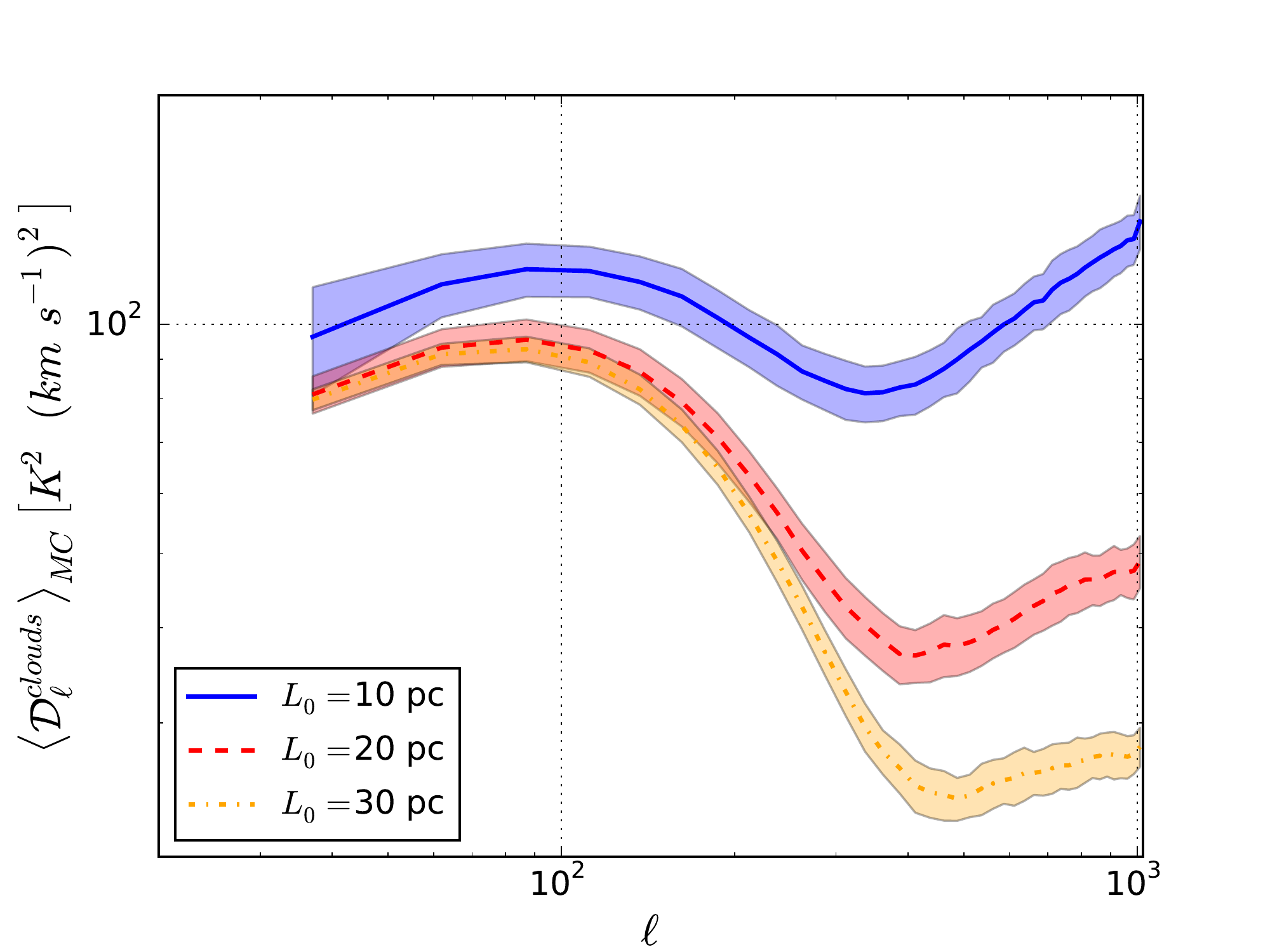}}
\subcaptionbox{}[\columnwidth]{\includegraphics[width=\columnwidth]{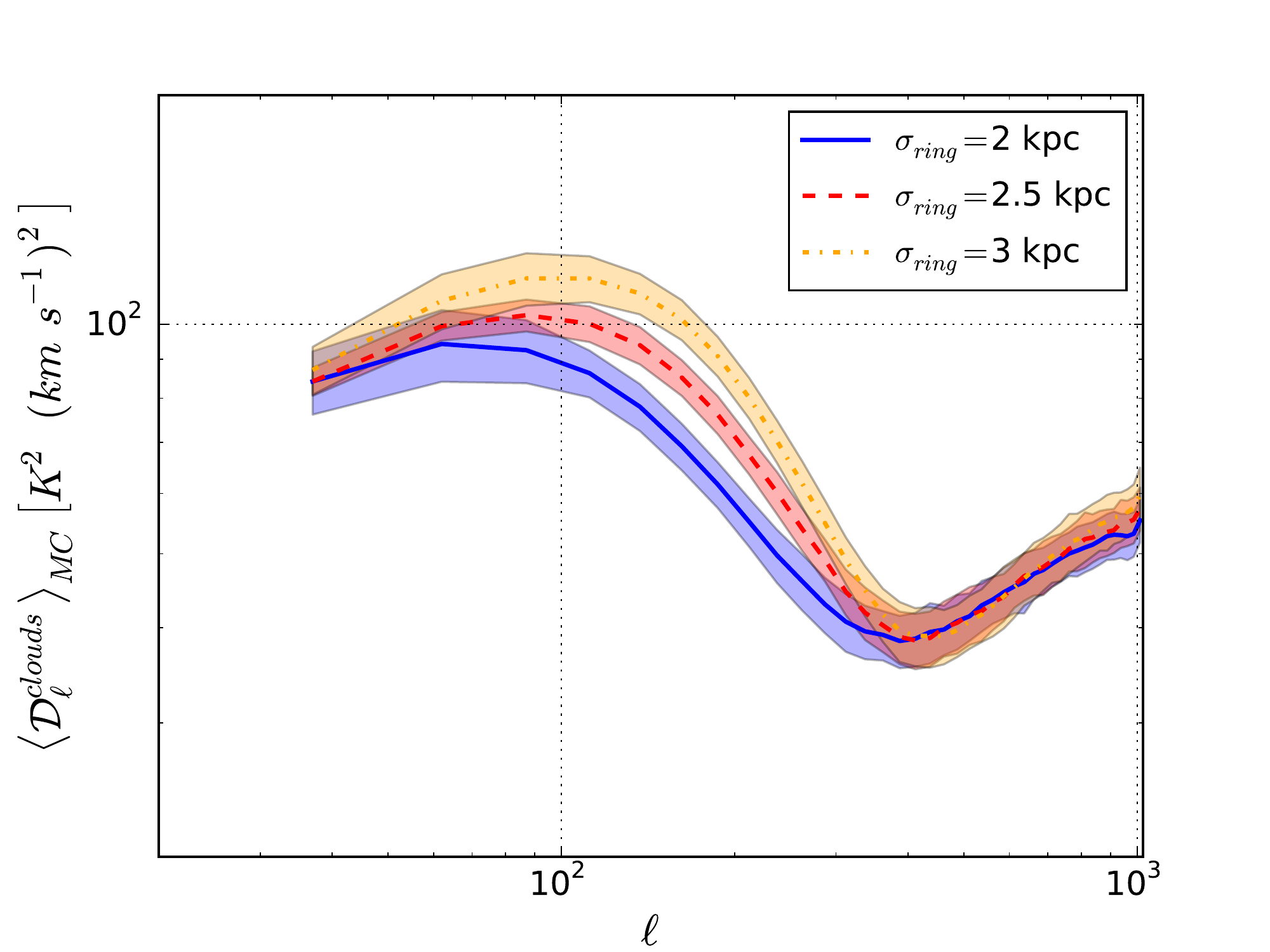}}\\
\caption{Angular power spectra of CO emission in the Galactic plane computed for 100 MC realizations of the \mcmole\ model assuming different values of its free parameters. The mean value of the simulation is shown by solid, dashed and dot-dashed  lines while the shaded area represents the measured variance of the realizations. The top row shows the case of an \axis\ geometry while the bottom panel displays results for a \spir\ geometry. Results obtained by varying the $L_0$ ($\sigma_{ring}$) parameters are shown on the left (right) column.}\label{fig:modspectra}
\end{figure*}

\subsection{Simulation results}
In \autoref{fig:mapsim} we show two typical realizations of maps of CO emission for the \axis\ and \spir\ geometries prior to any smoothing. As we are interested in the statistical properties of the CO emission, we report a few examples of the angular power spectrum $\mathcal{C}_{\ell}$ corresponding to different distributions of CO emission in \autoref{fig:modspectra}. In the ones shown subsequently the spectra are  $\mathcal{D}_{\ell}$ encoding a normalization factor $\mathcal{D}_{\ell} =\ell (\ell +1)\mathcal{C}_{\ell}/2\pi$. 

\begin{figure}
\includegraphics[width=\columnwidth]{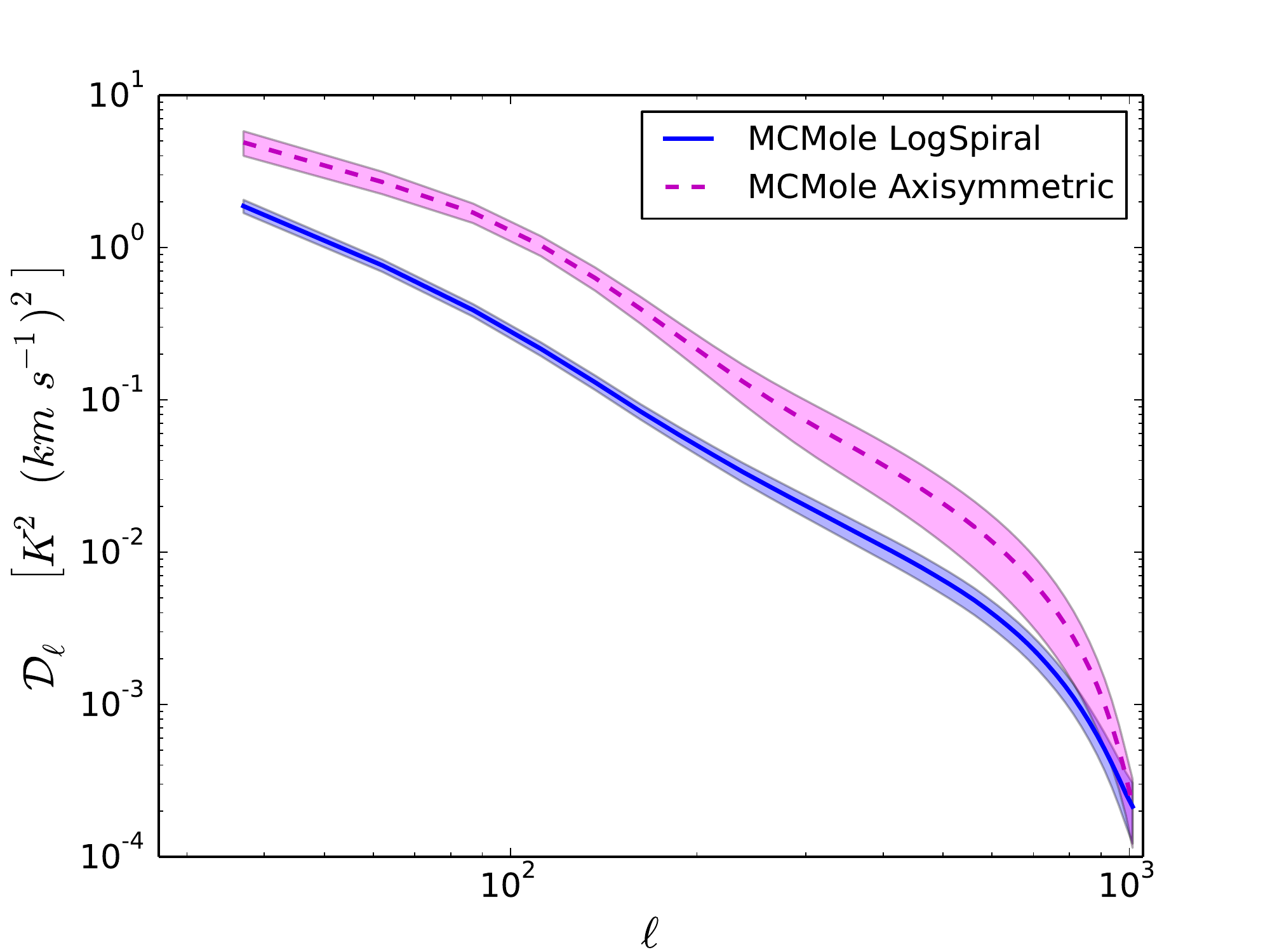}
\caption{Examples of the power spectra of CO emission at high Galactic latitudes ($|b|> 30$ deg) for \axis\ and \spir\ geometries. For both the geometries we assumed the best-fit values of the parameters describing the CO distribution presented in \autoref{sec:bestfit}.}\label{fig:HT}
\end{figure}
We can observe two main features in the morphology of the power spectrum: a bump around $\ell \sim 100 $ and a tail at higher $\ell$.  We interpret both of these features as the projection of the distribution of clouds from a reference frame off-centred (on the solar circle). \\*
The bump reflects the angular scale ($\sim 1$ deg) related to the clouds which have the most likely size, parametrized by the typical size parameter, $L_0$, and which are close to the observer. On the other hand, the tail at $\ell \gtrsim 600$ (i.e. the arcminute scale) is related to the distant clouds which lie in the diametrically opposite position with respect to the observer. This is the reason why the effect is shifted to smaller angular scales.
The $L_0$ and $\sigma_{ring}$ parameters modify the power spectrum in two different ways. For a given typical size, if the width of the molecular ring zone $\sigma_{ring}$ increases, the peak around $\ell \sim 100 $ shifts towards lower multipoles, i.e. larger angular scales, and its amplitude increases proportionally to $\sigma_{ring}$, see  for instance the bottom right panel in \autoref{fig:modspectra}.  
This can be interpreted as corresponding to the fact that the larger is $\sigma_{ring}$, the more likely it is to have clouds closer to the observer at the solar circle with a typical size given by $L_0$. 
On the other hand, if we choose different values for the size parameter (left panels in  \autoref{fig:modspectra}) the tail at small angular scales moves downwards and flattens as $L_0$ increases.   Vice versa, if we keep $L_0$ constant (\autoref{fig:modspectra} bottom right panel), all of the tails have the same amplitude and an $\ell^{2}$ dependency. In fact, if $L_0$ is small, the angular correlation of the simulated molecular clouds looks very similar to the one of \emph{point sources} which has \emph{Poissonian} behaviour.
Conversely if the typical size increases, the clouds become larger and they behave effectively as a coherent diffuse emission and less as point sources.\\*

Far from the Galactic plane, the shape of the power spectrum is very different. In \autoref{fig:HT} we show an example of the average power spectrum of 100 MC realizations of CO emission at high Galactic latitudes, i.e. $|b|>30$ deg, for both the \axis\ and \spir\ geometries. For this run we choose  the so-called best fit values for the $L_0$ and $\sigma_{ring}$ parameters discussed later in \autoref{sec:bestfit}. In addition to the different shape depending on the assumed geometry, one can notice a significant amplitude difference with respect to the power spectrum at low latitudes. Moreover, this is in contrast with the trend observed in the galactic plane, where the \spir\ geometry tends to predict a power spectrum of higher amplitude. In both cases, however, the model suppresses the emission in these areas, as shown in \autoref{fig:mapsim}. In the \spir\ case, the probability of finding clouds in regions in between spiral arms is further suppressed and could explain this feature. The emission is dominated by clouds relatively close to the observer for both geometries, and so the angular correlation is mostly significant at large angular scales (of the order of a degree or more) and  is damped rapidly at small angular scales.

\section{Comparison with Planck data}\label{sec:compars}
\subsection{Dataset}
The \pl\ collaboration released three different kinds of CO molecular line emission maps, described in \citet{2014A&A...571A..13P, plX2015}. We decided to focus our analysis on the so-called \emph{Type 1} CO maps which have been extracted exploiting differences in the spectral transmission of a given CO emission line in all of  the bolometer pairs relative to the same frequency channel. 
Despite being the noisiest set of maps, \emph{Type 1} are in fact the cleanest maps in terms of contamination coming from other frequency channels and astrophysical emissions. In addition, they have been obtained at the native resolution of the \pl\ frequency channels, and so allow full control of the effective beam window function for each map.\\* 
For this study we considered in particular the CO $1-0$ line, which has been observed in the 100 GHz channel of the HFI instrument. This channel is in fact the most sensitive to the CO emission in terms of signal-to-noise ratio (SNR) and the 1-0 line is also the one for which we have the most  detailed external astrophysical observations. 
However, the \pl\ frequency bands were designed to observe the CMB and foreground emissions which gently vary with frequency and, thus, they do not have the spectral resolution required to resolve accurately the CO line emission.
 To be more quantitative, the \pl\ spectral response at 100 GHz is roughly 3 GHz,  which corresponds to $\sim 8000\,  \mathrm{km\,s^{-1}}$, i.e. about 8 orders of magnitude larger than the CO rotational line width (which can be easily approximated as a Dirac delta). Therefore, the CO emission observed by \pl\  along each line of sight is integrated over the whole channel frequency band. Further details about the spectral response of the HFI instrument can be found in \citet{planck2013-spectral}.

\subsection{Observed CO angular power spectrum}\label{sec:co-ps}
\begin{figure}
\centering
\includegraphics[width=\columnwidth]{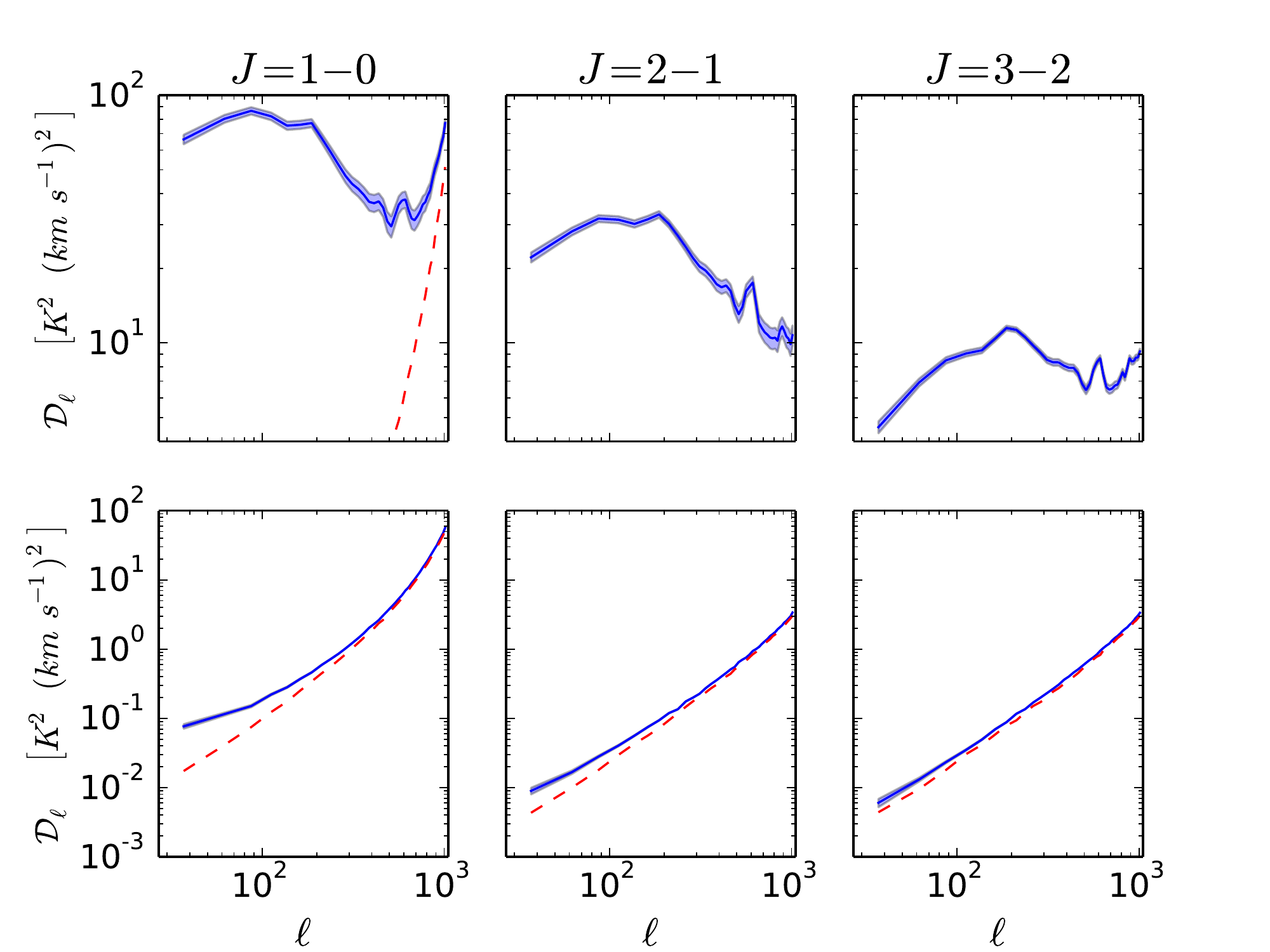}
\caption{CO $1-0$ angular power spectrum (blue solid) estimated from the \pl\ map at low (top) and high (bottom) Galactic latitudes. The shaded area shows the error bar due to the sample and noise variance. The expected noise level of the maps in the two regions is shown in (red dashed).}\label{fig:psgal1}
\end{figure}

Since one of the goals of this paper is to understand the properties of diffuse CO line emission, we computed the angular power spectrum of the \emph{Type 1} $1-0$ CO map to compare qualitatively the properties of our model with the single realization given by the emission in our Galaxy. We distinguish two regimes of comparison, low Galactic latitudes ($|b|\leq30\deg$) and high Galactic latitude ($|b|>30\deg$). While at low Galactic latitudes the signal is observed with high sensitivity, at high latitudes it is substantially affected by noise and by the fact that the emission in this region is faint due to its low density with respect to the Galactic disk.\\*
In \autoref{fig:psgal1} we show the angular power spectra  of the  first three CO rotational line maps observed by \pl\  as well as the expected noise level at both high and low Galactic latitudes computed using a pure power spectrum estimator \xpure\ \citep{2009PhRvD..79l3515G}.  This is a pseudo power spectrum method \citep{2002ApJ...567....2H} which corrects the so called E-to-B-modes leakage in the polarization field that arises in the presence of incomplete sky coverage \citep{PhysRevD.76.043001,2003PhRvD..67b3501B, PhysRevD.65.023505}. Although this feature is not strictly relevant for the analysis of this section, because we are considering the unpolarized component of the signal, it is important for the forecast presented in \autoref{sec:forecasts}. We estimated the noise as the mean of 100 MC Gaussian simulations based on the the diagonal pixel-pixel error covariance included in the \pl\ maps. One may notice how the noise has a level comparable to that of the CO power spectrum at high Galactic latitude. However, we note that the released \emph{Type 1} maps are obtained from the full mission data from Planck, and not from subsets of the data (e.g. using the so called half-rings or half-mission splits). Thus, it was not possible to test whether the observed flattening of the power spectrum at large angular scale is due to additional noise correlation not modelled by the Gaussian uncorrelated model discussed above. We notice that, if these maps were present, we could have had an estimate of this correlation using the noise given by the difference between the map auto-spectra and the noise-bias free signal obtained from the cross-spectra of the maps from data subsets. Since even for the 1-0 line, the noise becomes dominant on scales $\ell\approx 20$ we decided to limit the comparison at low Galactic latitude where the signal to noise ratio is very high. \\*
We note that in the following we considered the error bars on the power spectrum as coming from the gaussian part of the variance, i.e., following \cite{2002ApJ...567....2H}
\begin{equation}
\Delta \tilde{C}_{\ell} = \sqrt{\frac{2}{\nu}}(C_{\ell} + N_{\ell}) 
\end{equation}
\noindent
where $\nu$ is the number of degrees of freedom taking into account the finite number of modes going into the power spectrum calculation in each $\ell$ mode and the effective sky coverage. $N_{\ell}$ represents  the noise power spectrum and the $C_{\ell}$ is the theoretical model describing the CO angular power spectrum with the  tilde denoting  measured quantities. Because we do not know the true CO theoretical power spectrum we assumed that $C_{\ell} + N_{\ell} = \tilde{C}_{\ell}$. The gaussian approximation however underestimates the error bars. The CO field is in fact a highly non-gaussian field with mean different from  zero. As such, its variance should contain contributions coming from the expectation value of its 1 and 3 point function in the harmonic domain that are zero in the gaussian approximation. These terms are difficult to compute and we considered the gaussian approximation sufficient for the level of accuracy of this study.
\\*
As can be seen in \autoref{fig:psgal1}, all of the power spectra of CO emission at low Galactic latitudes have a broad peak around  the multipole $100\div 300$, i.e. at the $\approx1 \deg$ angular scale. The signal power starts decreasing up to $\ell \sim 600$ and then grows again at higher $\ell$ due to the Planck instrumental noise contamination.  
Such a broad peak suggests that there is a correlated angular scale along the Galactic plane. This can be understood with a quick order of magnitude estimate.   If we assume that most of the CO emission is localized at a distance of 4 kpc (in the molecular ring) and molecular clouds have a typical size of 30 pc, we find that each cloud subtends a $\sim 0.5$ deg area in the sky. This corresponds to a correlated scale in the power spectrum at an $\ell$ of the order of a few hundred but the detail of this scale depends on the width of the molecular ring zone.
\subsection{Galactic plane profile emission comparison}\label{sec:profile}
\begin{figure*}
\subcaptionbox{}[\columnwidth]{\includegraphics[width=\columnwidth]{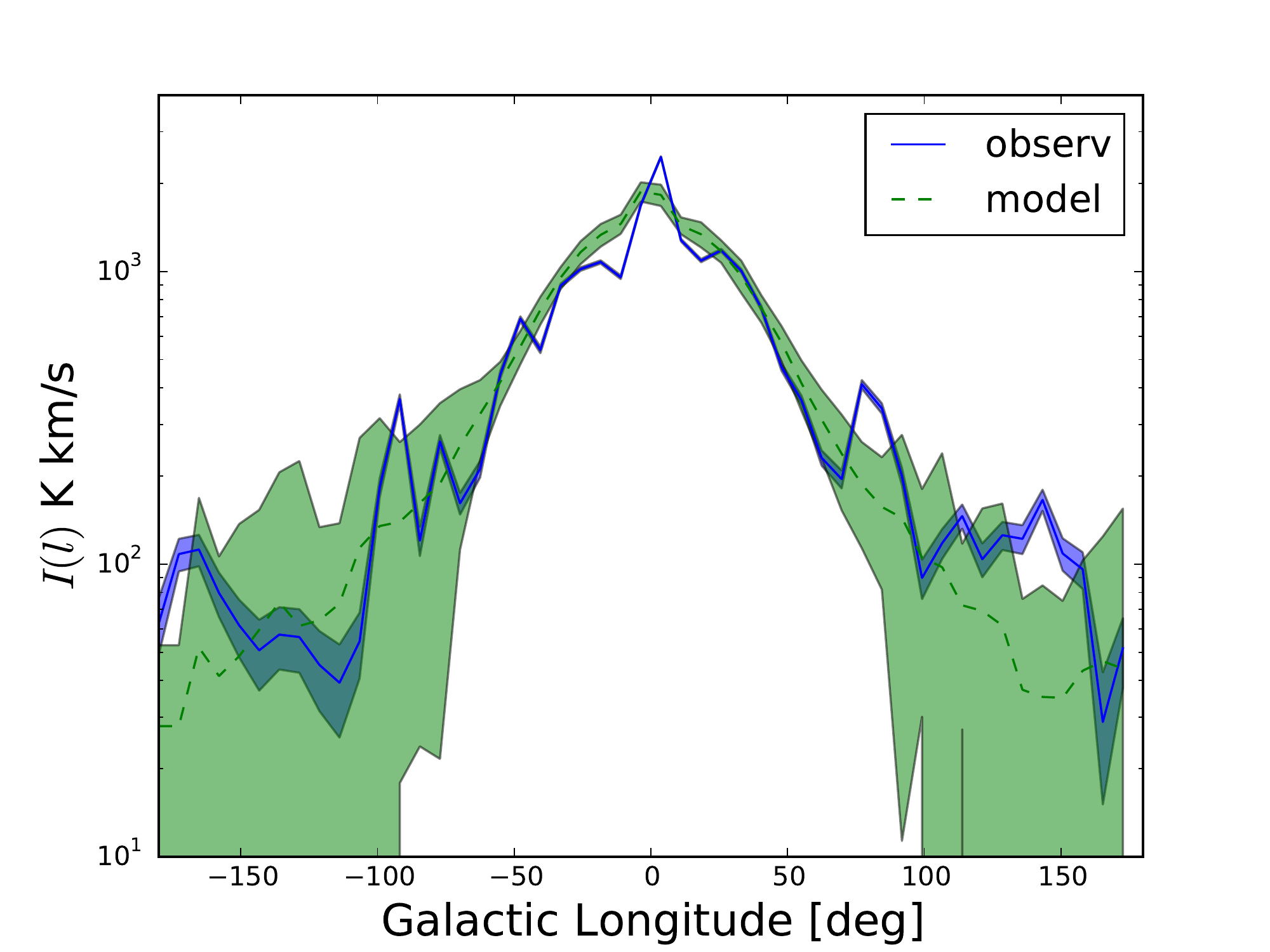}}
\subcaptionbox{}[\columnwidth]{\includegraphics[width=\columnwidth]{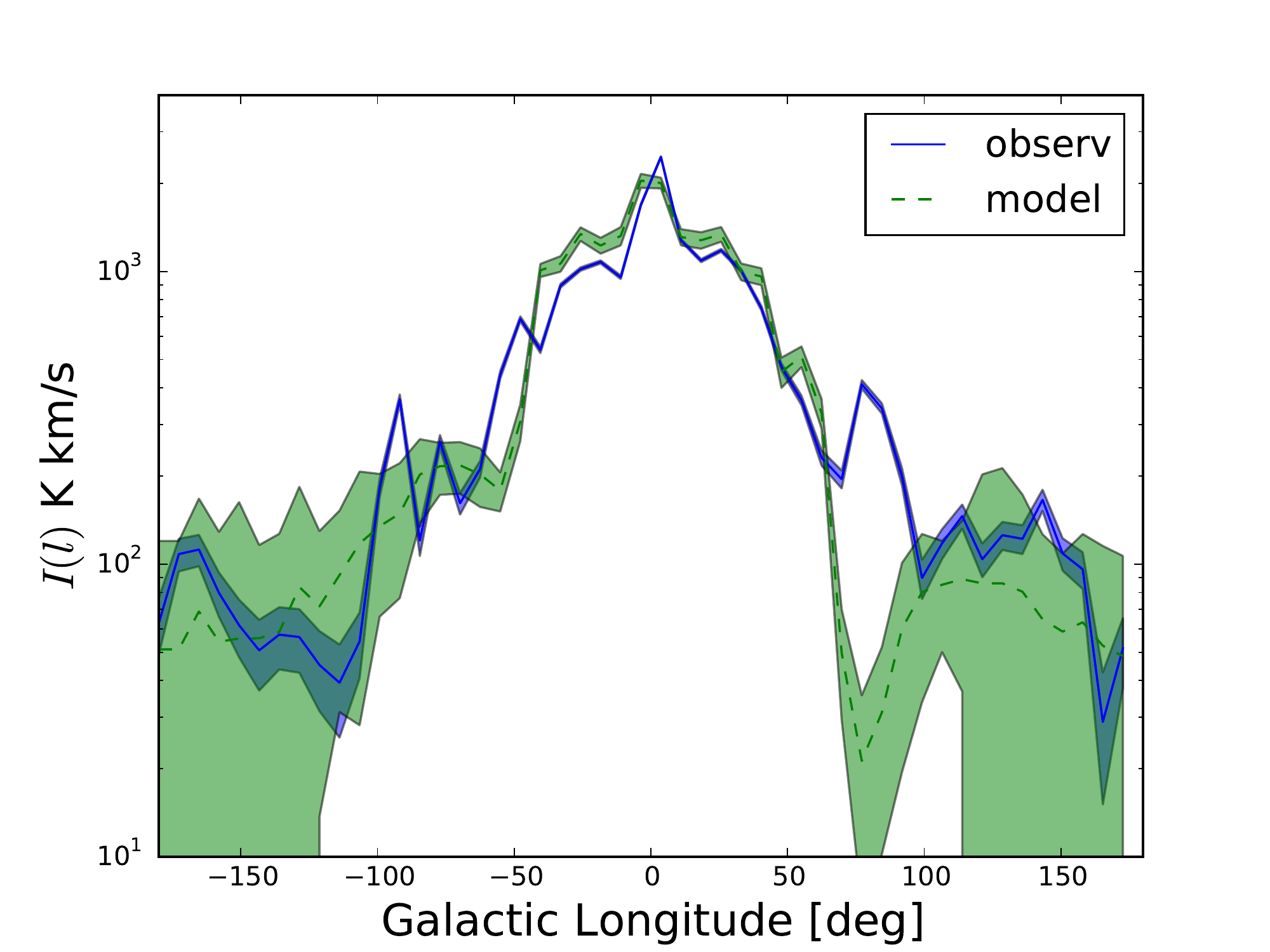}}
\caption{
Comparison of 100 MC realization of \mcmole\ simulated Galactic CO emission profiles (dashed green) with \pl\ observations (solid blue). The average profile integrals defined in \eqref{eq:modela} for the \axis\ (a) and \spir\ (b) geometries are  shown with dashed lines. The shaded area displays the standard deviation of all MCs in each longitude bin (green), or the noise level of \pl\ (solid blue) estimated from the \emph{Type 1} null map.} \label{fig:renorm}
\end{figure*}

As a first test we compared the profile of CO emission in the Galactic plane predicted by the model and the one observed in the data. Since we are mostly interested in a comparison as direct as possible with the \pl\ observed data, we convolved the \mcmole\ maps with a Gaussian beam of 10 arcmin FWHM, corresponding to the nominal resolution of the 100 GHz channel of HFI, prior to any further processing.\\*
In order to compare the data and the simulations, we constrained the total flux of the simulated CO maps with the one observed in the \pl\ data. This is necessary, otherwise the emission would be directly proportional to the number of clouds distributed in the simulated Galaxy.  Such a procedure also breaks possible parameter degeneracies with respect to the amplitude of the simulated power spectra (see next section). Following \citet{Bronfman1988}, we therefore computed the integrated flux of the emission along the two Galactic latitudes and longitudes $\left(l, b\right)$ defined as
\begin{align}
&I^{X}(l)=\int db I^{X}(l, b), \label{eq:modela}\\ 
&I_{tot}^{X}=\int dl db I^{X}(l,b), \label{eq:modelb}
\end{align}
where $X$ refers both to the \emph{model} and to  the \emph{observed} CO map. We then rescaled the simulated maps, dividing by the factor $f$ defined as:
\begin{equation}
f=\frac{I_{tot}^{observ}}{I_{tot}^{model}}.
\end{equation}
We estimated the integrals in \autoref{eq:modela} and \autoref{eq:modelb} by considering a narrow strip of Galactic latitudes within $\left[-2,2\right]$ degrees. We found that the value of $f$ is essentially independent of the width of the Galactic latitude strip used to compute the integrals because most of the emission comes from a very thin layer along the Galactic plane of amplitude $|b|\lesssim 2$ deg.\\*
In \autoref{fig:renorm} we show the comparison between  $I^{observ}(l)$ and the $I^{model}(l)$ as defined in \autoref{eq:modela} computed as the mean of 100 MC realizations of galaxies populated by molecular clouds  for both the \axis\ and \spir\ models as well as their typical standard deviation. In particular, we chose for these simulations the default parameters in \autoref{tab:params}. 

\noindent
The emission profiles are quite consistent in the regions from which most of the CO emission comes, i.e. in the inner Galaxy, the I and the IV quadrants (longitude in $\left[  -90 ,90 \right]$ deg\footnote{We stress that the definition of quadrants comes from the Galactic coordinates centred on the Sun. The I and IV quadrants are related to the inner Galaxy, while the II and the III ones look at its outer regions.}). On the contrary, the emission in the other two quadrants looks to be under-estimated but within the scatter of the simulations. In fact, the observed emissions in both the II and III quadrants come mainly from the closer and more  isolated system of clouds. These are actually more difficult to simulate because in that area (at Galactic longitudes $|l|>100$ deg) the presence of noise starts to be non-negligible (see shaded blue in \autoref{fig:renorm}). \\*
In addition, we note that the bump in the profile at $l \simeq 60 - 70$ deg, where we see a lack of power in both the \axis\ and \spir\ cases, corresponds to the complex region of \emph{Cygnus-X},  which contains the very well known X-ray source Cyg-X1, massive protostars and one of the most massive molecular clouds known, $3\times 10^6 \mathrm{M}_{\odot}$, 1.4 kpc distant from the Sun. 
Given the assumptions made in \autoref{sec:mod}, these  large and closer clouds are not easy to simulate with \mcmole\ especially where they are unlikely to be found, as in inter-spiral arm regions.
Despite of this, one can notice an overall qualitative better agreement with observations for the \spir\ model than for the \axis\ one. The latter reconstructs the global profile very well, but the former contains more peculiar features such us the central spike due to the Central Molecular Zone within the bar, or the complex of clouds at longitudes around $\sim -140,\,-80,\,120$ deg. We will perform a more detailed comparison of the two geometries in the following section and in \autoref{app:bestfit}.

\subsection{Constraining the \mcmole\ model with Planck data}\label{sec:bestfit}

After comparing the CO profile emission we checked whether the \mcmole\ model is capable of reproducing the characteristic shape of the Planck CO angular power spectrum. Given the knowledge we have on the shape of the Milky Way, we decided to adopt the \spir\ geometry as a baseline for this comparison, and to fix the parameters for the specific geometry to the values describing the shape of our Galaxy (see \autoref{sec:simproc}). For sake of completeness we reported the results of the same analysis adopting an \axis\ geometry in \autoref{app:bestfit}. \\*
We left  the typical cloud size $L_0$ and $\sigma_{ring} $ (the width of the molecular ring)  as free parameters of the model. While the former is directly linked to the observed angular size of the clouds,  the role of the second one is not trivial, especially if we adopt the more realistic 4 spiral arms distribution. Intuitively, it changes the probability of observing more clouds closer to the observer and affects more the amplitude of the power on the larger angular scales.\\*
We defined a large interval, reported in \autoref{tab:params}, where $L_0$ and $\sigma_{ring}$ are allowed to vary. Looking at the series of examples reported in \autoref{fig:modspectra} we can see that suitable parameter ranges which yield power spectra close to the \pl\ observations are $L_0=10 \div30$ pc and $\sigma_{ring}=2 \div 3$ kpc. It is interesting to note that these are in agreement with estimates available in the literature (see e.g. \citep{Ellsworth-Bowers2015,Roman-Duval2010}).\\*
We then identified a set of values within the intervals just mentioned for which we computed the expected theoretical power spectrum of the specific model. Each theoretical model is defined as the mean of the angular power spectrum of 100 MC realizations of the model computed with \xpure. For each realization of CO distribution we rescaled the total flux following the procedure outlined in the previous section before computing its power spectrum. \\*
Once the expected angular power spectra for each point of the parameter domain had been computed, we built the hyper-surface $\mathcal{F}(\ell; \sigma_{ring}, L_0)$ which for a given set of values $(\sigma_{ring}, L_0)$ retrieved the model power spectrum, by  interpolating it from its value at the closest grid points using splines. We checked that alternative interpolation methods did not impact significantly our results. We then computed the best-fit parameters of the \mcmole\ model by performing a $\chi^{2}$ minimization with the Planck CO power spectrum data. For this procedure we introduced a further global normalization parameter $A_{CO}$ to take account of the \pl\ bandpass effects or other possible miscalibration of the model. These might come either from variations from the scaling laws employed in the model (that are thus not captured by the total flux normalization described earlier), or calibration differences between the \pl\ data and the surveys used to derive the scaling laws themselves. The bandpass effects tend to decrease the overall amplitude of the simulated signal because each line gets diluted over the width of the \pl\ frequency band.\\*
Since the theoretical model has been estimated from Monte Carlo simulations, we added linearly to the sample variance error of the \pl\ data an additional uncertainty budget corresponding to the uncertainty of the mean theoretical power spectrum estimated from MC. We note that when we compute the numerator of the $f$ rescaling factor, we include not only the real flux coming from the CO lines but also an instrumental noise contribution. We therefore estimated the expected noise contribution to $f$ by computing the integral of \autoref{eq:modelb} on the \pl\ error map and found it to be equal to 10\%. We propagated this multiplicative uncertainty to the power spectrum level, rescaling the mean theoretical MC error bars by the square of this factor.

\begin{figure*}
\begin{flushleft}
\subcaptionbox{}[\columnwidth]{\includegraphics[scale=.5,trim= 0 0 0 0cm,clip=true]{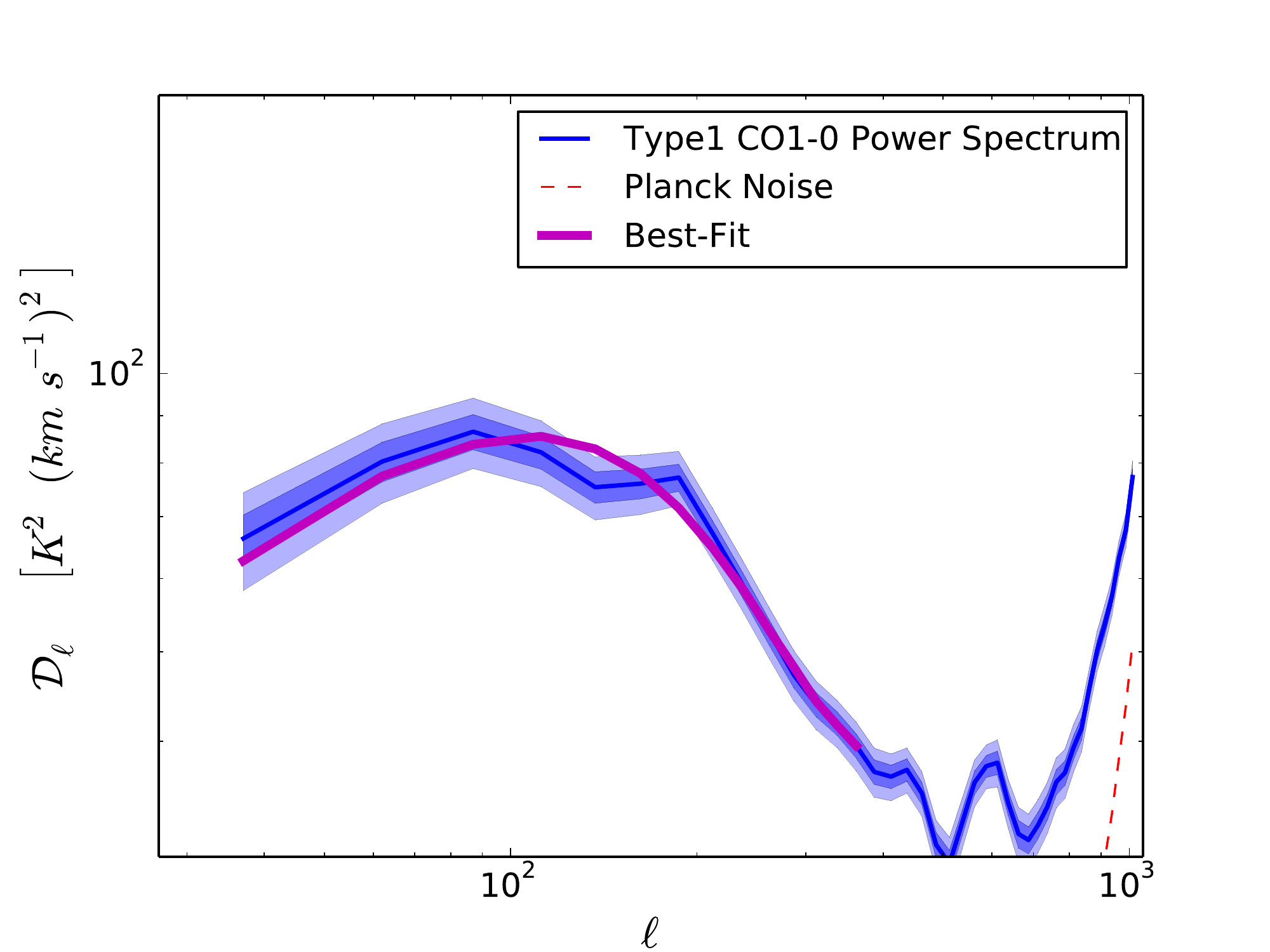}}
\hspace{.6cm}
\subcaptionbox{}[\columnwidth]{\includegraphics[scale=.5,trim=  2.4cm 0 0 0cm,clip=true]{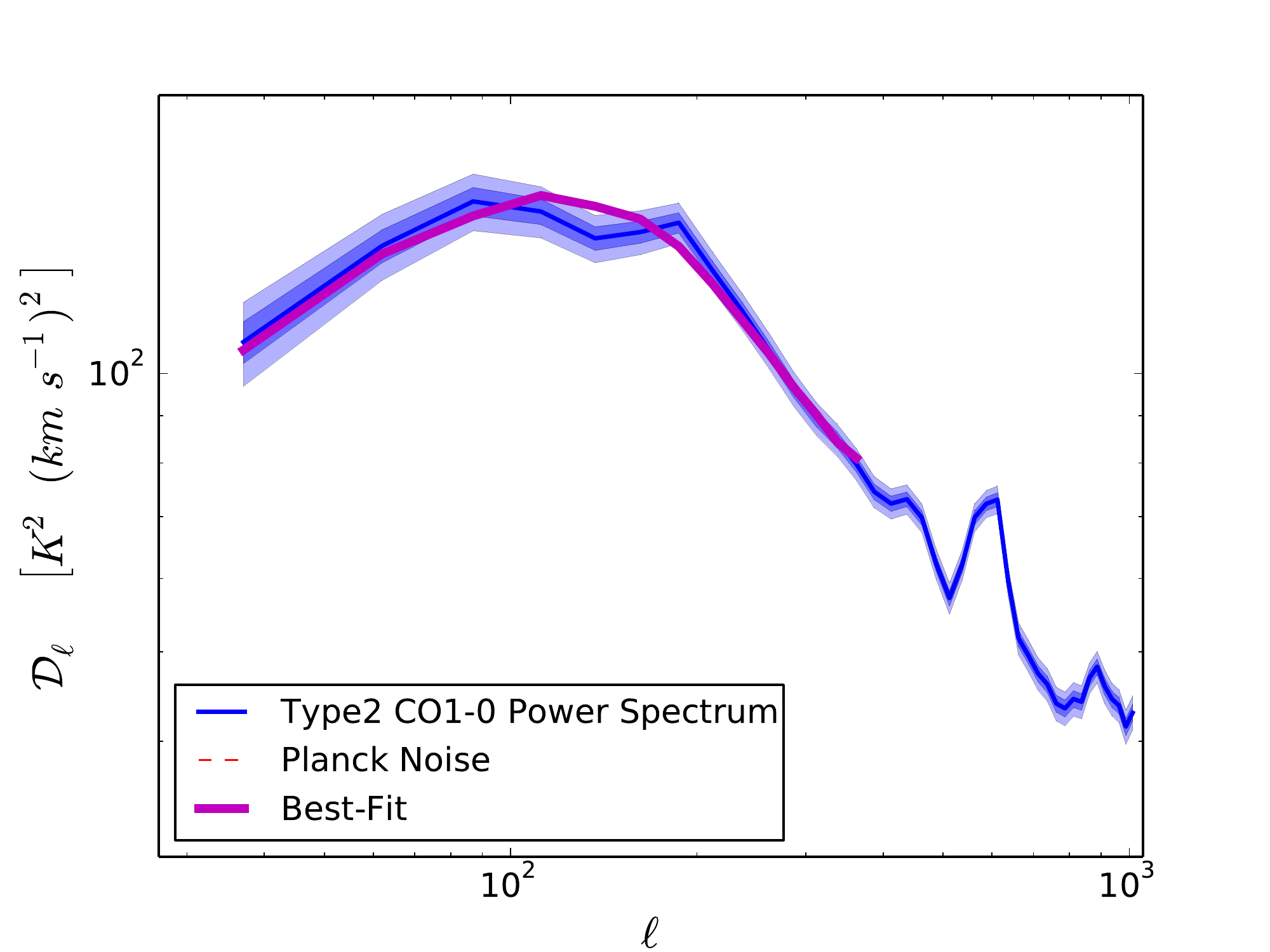}}

\subcaptionbox{}[\columnwidth]{\includegraphics[scale=.4]{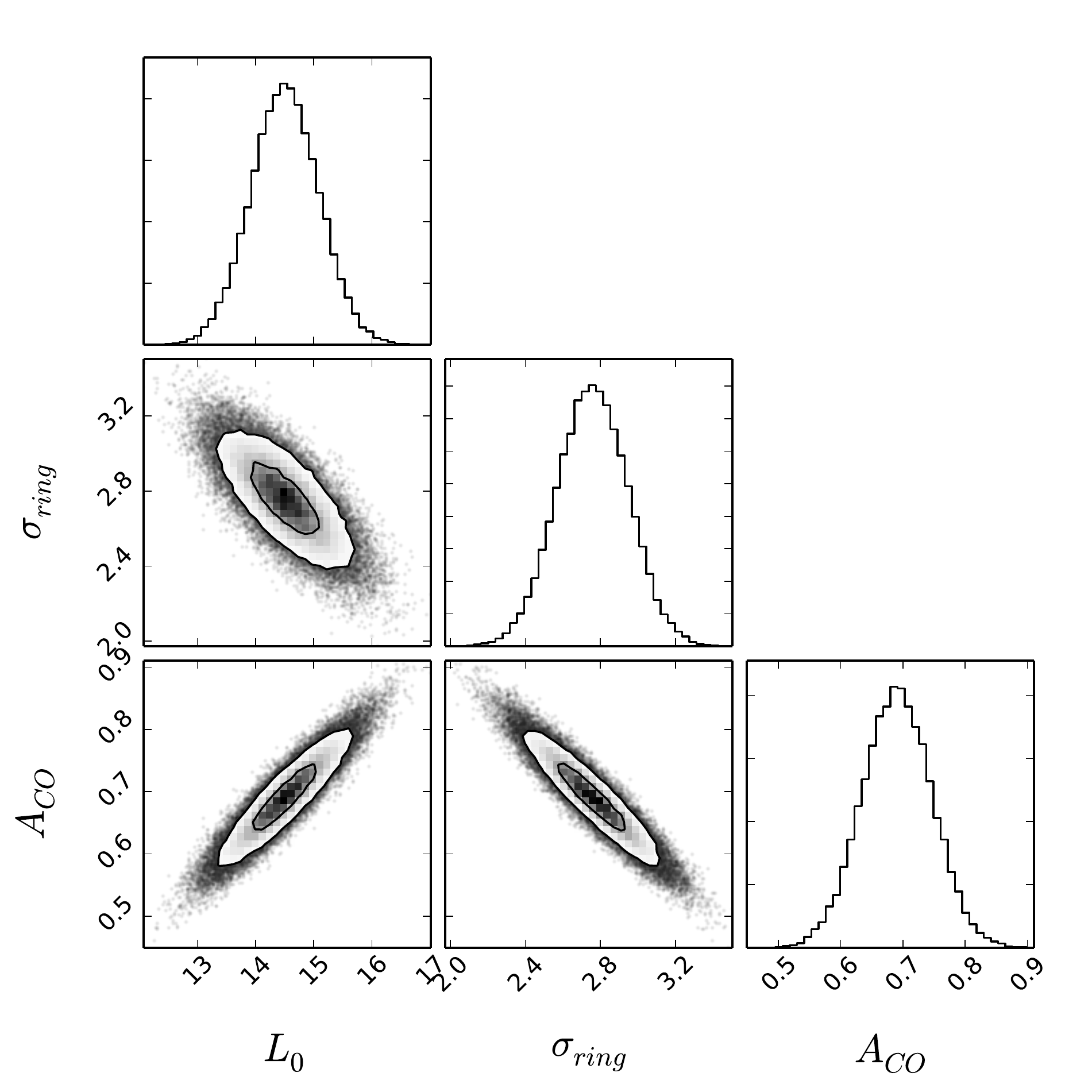}}
\subcaptionbox{}[\columnwidth]{\includegraphics[scale=.4]{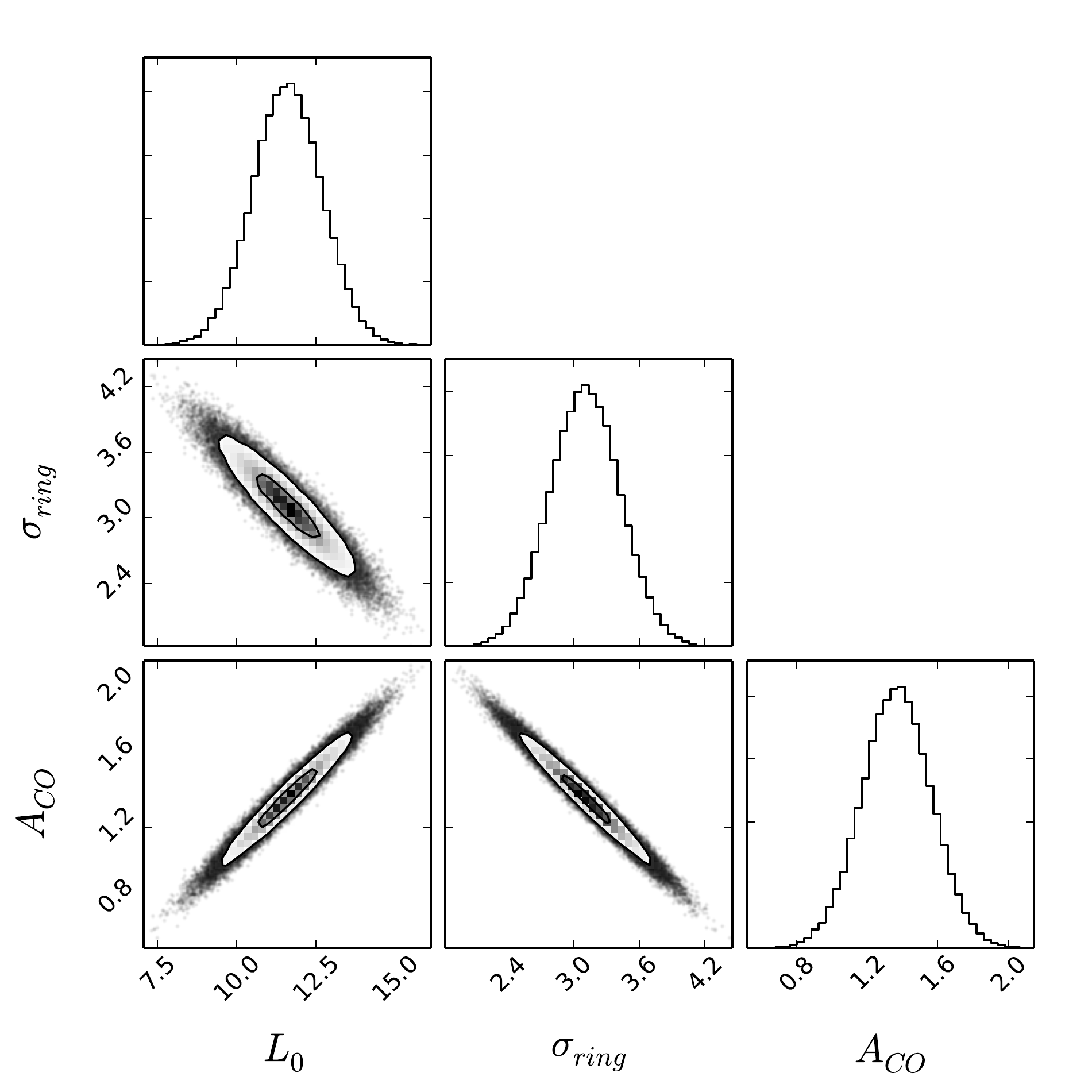}}\\
\caption{\emph{Top panels}: angular power spectra (solid thin blue) of the \pl\ \emph{Type 1} (left) and \emph{Type 2} (right)  maps. The shaded area correspond to the  $1 \sigma $ (dark blue) and $2 \sigma$ (light blue) error bars including statistical and systematic uncertainties. The \mcmole\ model CO angular power spectrum assuming the best-fit parameters of \autoref{eq:bestfit} is shown in thick solid magenta. The \pl\ noise power spectrum is shown in red dashed, in the top right panel the noise level is about one order of magnitude smaller than the one in the top left panel. \emph{Bottom panels}: best-fit parameters of the \mcmole\ model describing the \pl\ CO angular power spectrum and their correlations.}\label{fig:bestfit}

\end{flushleft}
\end{figure*}
\begin{table*}
\large 
\begin{tabular}{c c c c c c c c}
\toprule 
 & $L_0$ [pc]& $\sigma_{ring}$ [kpc] & $A_{CO}$& $\tilde{\chi}^2$ & \emph{dof} & $p$-value & $\rho_{L\sigma} $\\
\midrule
Type 1 & $14.50 \pm 0.58$ & $ 2.76 \pm 0.19$ & $0.69 \pm 0.06$ & $1.48$ &  11 & 0.13 &  0.74\\
Type 2 & $11.59 \pm 1.09$ & $3.11 \pm 0.32$  & $1.37 \pm 0.19$ & 1.95 & 11 & 0.03&0.92 \\
\bottomrule
\end{tabular}
\caption{Summary table of best fit parameters obtained using the two different \pl\ CO maps. }\label{tab:bestfit_params}
\end{table*}
\noindent
We limit the range of angular scales involved in the fit to $\ell\leq400$ in order  to avoid the regions that display an unusual bump at scales of around $\ell\approx 500$ that is not captured by any realization of our model (see next section). The best-fit parameters are reported in \autoref{tab:bestfit_params}
\begin{align}
&L_0 =  14.50 \pm 0.58\, \mathrm{pc}, \nonumber \\
&\sigma_{ring}= 2.76 \pm 0.19\, \mathrm{Kpc}, \label{eq:bestfit}\\
&A_{CO} = 0.69 \pm 0.06 . \nonumber 
\end{align}
The values are within the ranges expected from the literature. As can be seen in  \autoref{fig:bestfit} the power spectrum corresponding to the model with the best fit parameters, describes the \pl\ data reasonably well. The minimum $\chi^{2}$ obtained by the minimization process gives $1.48$ that corresponds to a $p$-value of $13 \%$. We note, however,that all of the parameters are highly correlated. This is somewhat expected as the larger is $\sigma_{ring}$, the closer the clouds get to the observer placed in the solar circle. This effect can be compensated by an overall decrease of the typical size of the molecular cloud as shown in \autoref{fig:modspectra}(d).\\*

Finally, we note that $A_{CO}\lesssim 1$ suggests that, despite the rescaling procedure constraining quite well the overall power spectrum amplitude, the spatial distribution seems to be more complex than the one implemented in the model. This might partially be explained by the fact that we do not model explicitly any realistic bandpass effect of the \pl\ channel or the finite width of the CO line. Additional sources of signal overestimation could be residual contamination of $^{13}CO$ 1-0 line or thermal dust in the map or variations of the emissivity profile in \autoref{eq:emprof}.

\subsection{Consistency checks on other maps}
The \pl\ collaboration released multiple CO maps extracted using different component separation procedures. 
We can test the stability of our results by using CO maps derived with these different approaches, in particular the so-called \emph{Type 2} maps. These have been produced exploiting the intensity maps of several frequencies (\emph{multi-channel} approach) to separate the CO emission from the astrophysical and CMB signal \citep{2014A&A...571A..13P}. The maps are smoothed at a common resolution of $15$ arcmin  and have better S/N ratio than the \emph{Type 1} ones. However, the CO is extracted by assuming several simplifications which may leak into contamination due to foreground residuals and systematics, as explained in \citet[section 5.5.3]{plX2015}.\\*
We repeated the procedure outlined in \autoref{sec:profile} and \autoref{sec:bestfit} using the \emph{Type 2} $1-0$ map. The values of the best fit parameters are summarized in \autoref{tab:bestfit_params} and we show in \autoref{fig:bestfit} the best-fit model power spectrum together with the power spectrum of \emph{Type 2} map data. We found that the values of $A_{CO}$ obtained for \emph{Type 2} are inconsistent with the one obtained for the \emph{Type 1} maps. However, this discrepancy is consistent with the overall inter calibration difference between the two maps reported in \citet{plX2015}.
Such differences are mainly related to a combination of bandpass uncertainties in the \pl\ observations and presence of a mixture of ${}^{12} CO$ and ${}^{13} CO$  (emitted at $110$ GHz) lines for the \emph{Type 1 maps}. 
While $\sigma_{ring}$ is consistent between the two maps, the \emph{Type 2} $L_{0}$ parameters are in slight tension at $2.7\sigma$ level. The overall correlation of the parameters is increased and the overall agreement between data and the \mcmole\ mode is reduced although it remains acceptable. We cannot exclude however that this is a sign of additional systematic contamination in the \emph{Type 2} maps.\\*

The \pl\ collaboration provided maps of the 2-1 line for both of the methods and we could use our model to constrain the relative amplitudes of the lines, while fixing the parameter of the cloud distribution. However, such analysis is challenging and might be biased by the presence of variations of local physical properties of the clouds (opacity and temperature) or by the red or blue shift of the CO line within the \pl\ bandpass induced by the motion of the clouds themselves \citep{plX2015}. For these reasons, we decided to restrict our analysis only to the CO $1-0$ line, since it is the one for which the observational data are more robust. 
 
We finally note that the observed angular power spectra of the \pl\ maps display an oscillatory behaviour at a scale of $\ell\geq 400$ with a clear peak at around $\ell\approx 500$. The fact that this feature is present in all of the lines and for all of the CO extraction methods means that it can reasonably be considered as a meaningful physical signature. Because a single cloud population produces an angular power spectrum with a characteristic peak scale, we speculate that this could be the signature of the presence of an additional cloud population with a different typical size or location. We however decided to leave the investigation of this feature for a future work.

\subsection{Comparison with data at high Galactic latitudes}
\label{subsec:hglcomparison}
\begin{figure}
\includegraphics[width=\columnwidth]{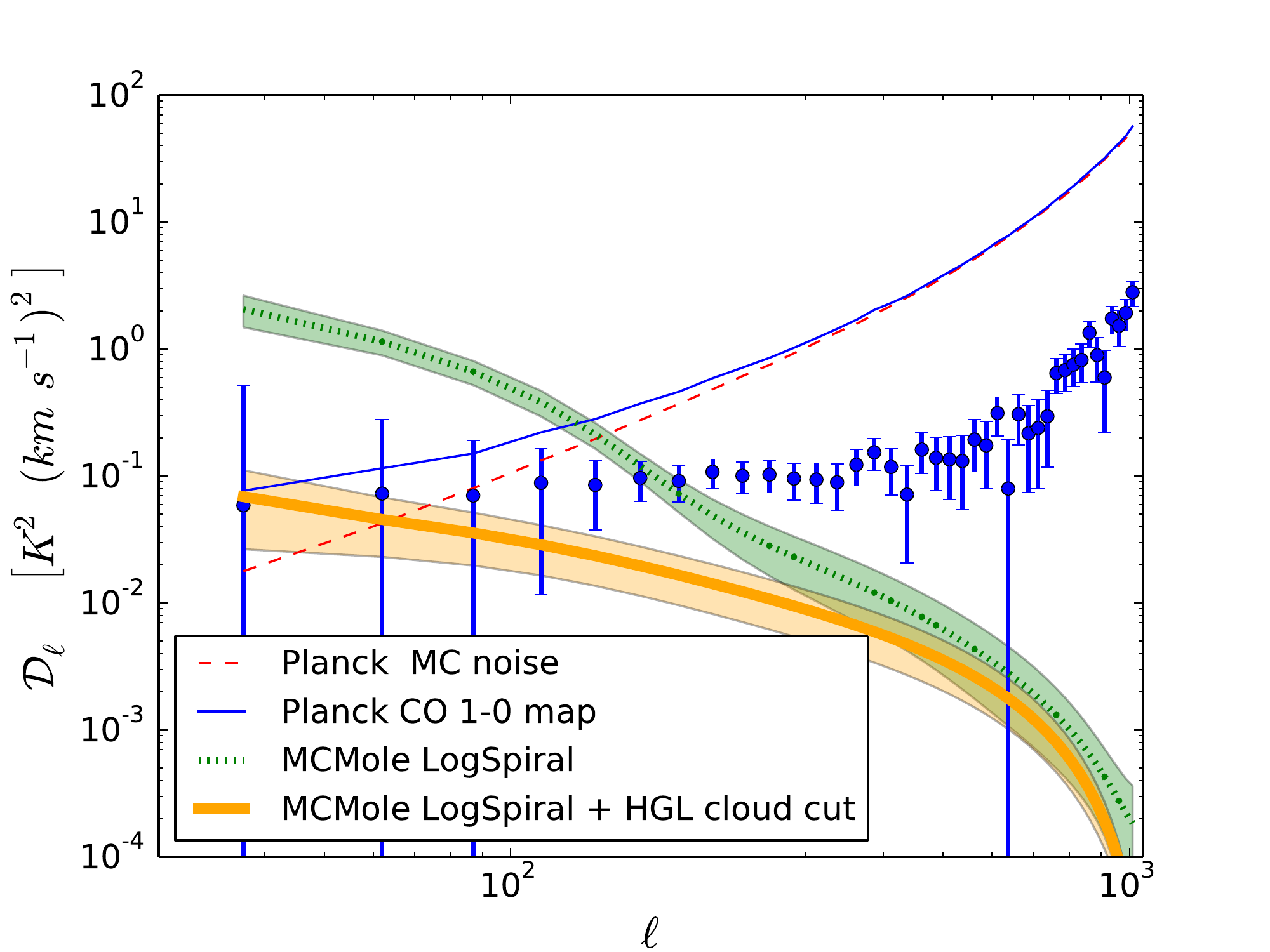}
\caption{ CO 1-0 power spectrum at high Galactic latitudes   of the \spir\ \mcmole\ model (dotted green), using the parameters in \eqref{eq:bestfit}. (thick solid orange) We show the power spectrum for the  \spir\ geometry,  the same parameters in \eqref{eq:bestfit} and with  the HGL-cut of clouds at  $|b|> 30 $ deg whose flux exceeds the \pl\ noise. The \pl\ \emph{Type 1} CO power spectrum before and after noise bias subtraction is shown  with the blue solid line and filled circles respectively;  the error bars account for both \pl\ data  statistical uncertainties and   systematics from the \mcmole\ simulations. The noise bias is shown with the dashed red line.}\label{fig:bestfit_ht}
\end{figure}

In \autoref{fig:bestfit_ht}, we compare the \pl\ CO 1-0 power spectrum at high Galactic latitudes with the average power spectrum of 100 MC realizations of the \mcmole\ model for the same region of the sky. We assumed for these runs the best-fit values of the $L_0,\, \sigma_{ring}$ parameters reported in \autoref{eq:bestfit} and a \spir\ distribution. Because the \pl\ maps at these latitudes are dominated by noise, we subtracted our MC estimates of the noise bias data power spectrum so as to have a better estimate of the signal (blue circles).  

As can be observed in \autoref{fig:bestfit_ht},  some discrepancy arises when comparing the power spectrum expected from the simulation of \spir\ \mcmole\ at high Galactic latitudes with the noise debiased \pl\ data. This is rather  expected because the model has larger uncertainties at high Galactic latitudes than in the Galactic mid-plane (where the best-fit parameters are constrained) given the lack of high sensitivity data. The discrepancy seems to point to an overestimation of the vertical profile parameters $\sigma_{z,0}$ and $h_R$ (see \autoref{eq:zfunc}) which gives  a higher number of clouds close to the observer at high latitude. 
However, we also point out that, as explained in \autoref{sec:co-ps}, the error bars in \autoref{fig:bestfit_ht} might be underestimated especially at the largest angular scales where we are signal dominated. Therefore, discrepancies of  order $\approx 3\sigma$ do not seem unlikely. 
Since we are mainly interested in using the model to forecast the impact of unresolved CO emission far from the Galactic plane ($|b|>30$), we investigated whether removing the few high Galactic latitude clouds in the simulation that appear close to the observer would improve the agreement with the \pl\ data. All of these clouds have, in fact, a flux exceeding the \pl\ CO map noise in the same sky area and they should have already been detected in real data. We will refer to this specific choice of cut as the High Galactic Latitudes (HGL) cut in the following.  The power spectrum of the \mcmole\ simulated maps obtained after the application of the  HGL cut is shown in \autoref{fig:bestfit_ht}.
 We found that the model calibrated at low latitudes and after the application of the HGL-cut agrees very well with the data on the angular scales where the signal slightly dominates, i.e. $\ell\lesssim 80$. We could not extend the comparison to smaller angular scales because the data  become noise dominated and the residual increase of power observed on the power spectrum is dominated by a noise bias residual.

\section{Polarization Forecasts}\label{sec:forecasts}

As noted in \autoref{sec:intro}, CO lines are polarized and could contaminate sensitive CMB polarization measurements together with other polarized Galactic emission (synchrotron and the thermal dust) at sub-millimeter wavelengths.
Future experiments will preferentially perform observations at intermediate and high Galactic latitudes, to minimize contamination from  strong Galactic emissions close to the plane. Since CO data at high Galactic latitudes are not sensitive enough to perform accurate studies of this emission, we provide two complementary estimates of the possible contamination from its polarized counterpart to the CMB B-mode power spectrum in this sky region.

\subsection{Data-based order of magnitude estimate}\label{subsec:rescalingspectra}
Starting from the measured \pl\ power spectrum at low Galactic latitudes, one can extrapolate a very conservative value of the CO power spectrum at higher latitudes. Assuming that all of the variance observed in the high Galactic latitude region is distributed among the angular scales in the same way as in the Galactic plane, we can write
\begin{equation}
\mathcal{C}_{\ell}  ^{high, CO}= \mathcal{C}_{\ell}  ^{Gal} \frac{\mathrm{var}(high)}{\mathrm{var}(Gal)} .
\end{equation}

This is a somewhat conservative assumption because we know that the bulk of the CO line emission is  concentrated close to the Galactic disk and also because it assumes that the \pl\ noise at high Galactic latitudes is diffuse CO emission. The variance of the \pl\ CO map is $0.3 \, \mathrm{K^2 (km \,s^{-1})^2 }$, at $|b|>30$ deg, while for $|b|<30$ deg we get a variance of $193.5$ $\mathrm{K^2 \,(km\, s^{-1})^2 }$. 
Taking $1\% $ as the polarization fraction, $p_{CO}$, of the CO emission and an equal power in E and B-modes of polarized CO, we can convert $\mathcal{C}_{\ell}^{CO high}$  into its B-mode counterpart as $\mathcal{C}_{\ell} ^{CO high, EE}=\mathcal{C}_{\ell} ^{CO high, BB}= \mathcal{C}_{\ell}^{high, CO}p_{CO}^{2}/2$. We then apply the conversion factors of \citet{2014A&A...571A..13P} to convert the CO power spectrum into thermodynamic units (from $\mathrm{K_{RJ} km s^{-1}}$ to $\mu \mathrm{K}$). We can compare $\mathcal{C}_{\ell} ^{CO high, BB}$ to the amplitude of  equivalent cosmological CMB inflationary B-modes with tensor-to-scalar ratio $r=1$ at $\ell={80}$.  In terms of $\mathcal{D}^{BB}_{\ell}$, this  is equal to $\sim 6.67 \times 10^{-2} \mu \mathrm{K}^2$ for a fiducial Planck 2015 cosmology. We found that the amplitude of the extrapolated CO B-mode power spectrum is equal to a primordial B-mode signal having $r_{CO} =0.025$. 

\subsection{Simulation estimate}\label{subsec:pol-sim}
In order to verify and refine the estimate given in the previous Section, we used the model presented in \autoref{sec:mod} to infer the level of contamination from unresolved polarized CO emission.
For doing this, we first set the free parameter of the \mcmole\ model to the best-fit value derived in \autoref{eq:bestfit}. \\*
From the total unpolarized emission in each sky pixel of the simulation, $I^{CO}$ we can then extract its linearly polarized part by taking into account  the global properties of the Galactic magnetic field. Following  \cite{2013A&A...553A..96D, tassis2015} the Q and U Stokes parameters of each CO cloud can be related to the unpolarized emission as

\begin{align}
Q(\hat{n})^{CO}&=p_{CO}\,g_d(\hat{n})I(\hat{n})^{CO}\, cos(2 \psi(\hat{n})),\label{eq:q}\\
U(\hat{n})^{CO}&=p_{CO}\,g_d(\hat{n})I(\hat{n})^{CO}\, sin(2 \psi(\hat{n})).\label{eq:u}.
\end{align} 
\noindent
 where $p_{CO}$ is the intrinsic polarization fraction of the  CO lines, while $g_d$ is the geometric depolarization factor which accounts for the induced depolarization of the light when integrated along the line of sight. The polarization angle $\psi$ describes the orientation of the polarization vector and, for the specific case of Zeeman emission, it is related to the orientation of the component of the Galactic magnetic field orthogonal to the line of sight $B_{\bot}$.  Following the findings of \cite{Greaves1999}, we adopted a conservative choice of a constant $p_{CO}=1 \%$ for each molecular cloud of the simulation. Because the polarized emission in molecular clouds is correlated with the polarized dust emission \citep[see, e.g.]{2012ARA&A..50...29C}, we used the $g_d$ and $\psi$ templates for the Galactic dust emission available in the public release of the Planck Sky  Model suite\footnote{\url{http://www.apc.univ-paris7.fr/~delabrou/PSM/psm.html}} \citep{2013A&A...553A..96D}. These have been derived from 3D simulations of the Galactic magnetic field (including both a coherent and a turbulent component) and data of the WMAP satellite. \\*
Since we assumed a constant polarization fraction, the geometrical depolarization effectively induces a change in the polarization fraction as a function of Galactic latitudes decreasing it when moving away from the poles. 
This effect has already been confirmed by \pl\ observations \citep{plX2015} of thermal dust, whose polarization fraction increases at high latitudes.\\*
In order to forecast the contamination of unresolved CO polarized emission alone, we apply the HGL-cut as described in \autoref{subsec:hglcomparison} to each realization of the model for consistency. 

\begin{figure}
\includegraphics[width=\columnwidth]{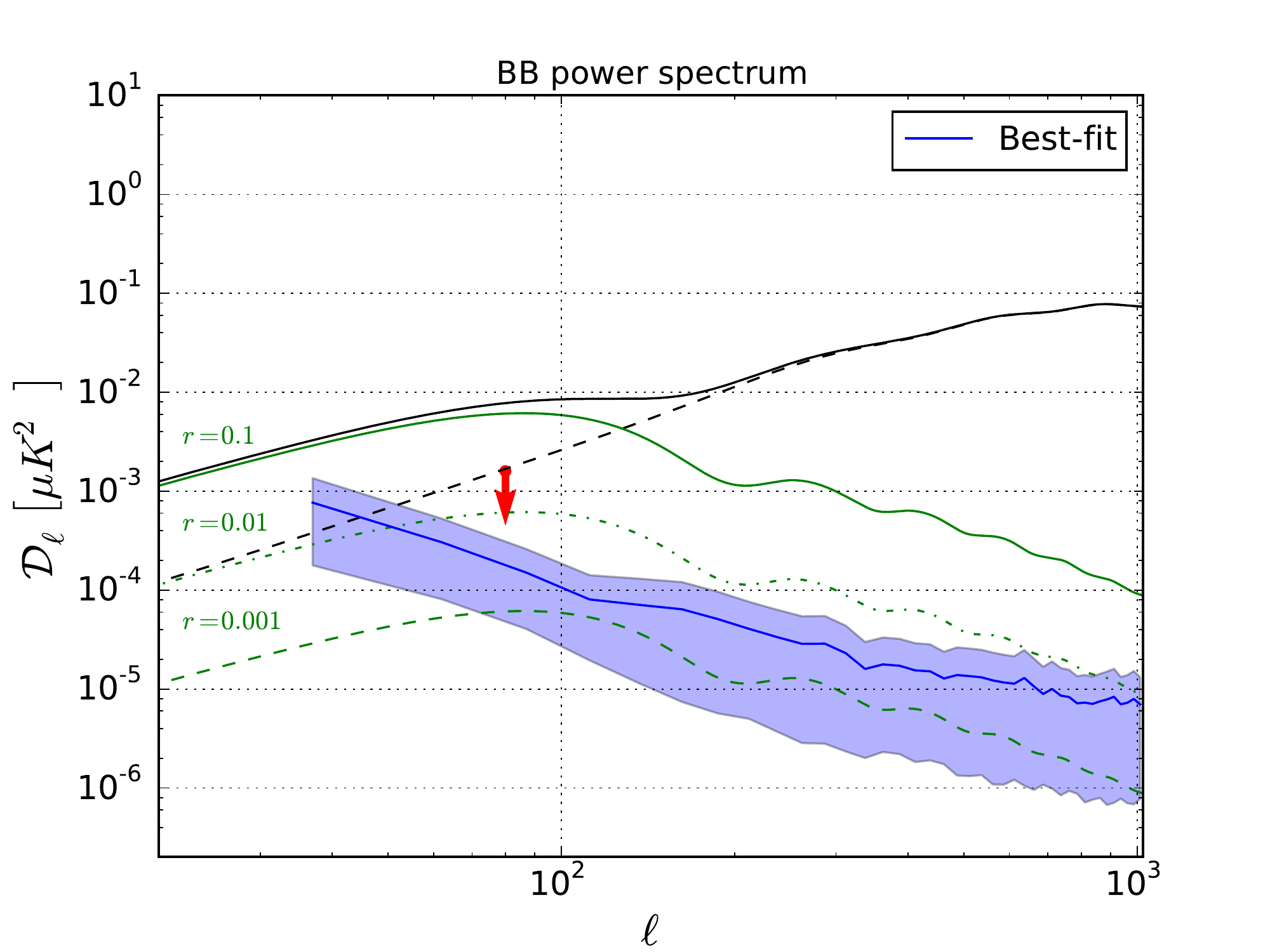}
\caption{ B-mode power spectra of polarized CO emission lines at high Galactic latitudes estimated using the best-fit parameters of the \spir\  \mcmole\ model (see \autoref{eq:bestfit}). The expected Planck 2015 $\Lambda$CDM cosmological signal including the gravitational lensing contribution is shown in black. Potential contributions from  inflationary B-modes for tensor-to-scalar ratios of  $r=0.1,0.01,0.001$ are shown with solid, dot-dashed and dashed green lines respectively. The red arrow indicates the upper limits obtained  in \autoref{subsec:rescalingspectra} }
\label{fig:pol}
\end{figure}

Once the $Q^{CO}$ and $U^{CO}$ maps have been produced, we computed the angular power spectrum using \xpure.\\*
In \autoref{fig:pol} we show the mean and standard deviation of the B-mode  polarization power spectrum extracted from 100 MC realizations of the CO emission following the procedure just outlined. Even though in \autoref{sec:bestfit} we showed that our model tends to slightly overestimate the normalization of the power spectrum, we decided not to apply the best-fit amplitude $A_{CO}$ to the amplitude of the B-mode power spectrum in order to provide the most conservative estimates of the signal. \\*
As could be seen  from \autoref{fig:pol}, there is a significant dispersion compared to the results of the MC simulations at low Galactic latitudes (see \autoref{fig:modspectra}). This simply reflects the fact that the observations, and hence our model, do not favour the presence of molecular clouds at high Galactic latitudes. Therefore their number can vary significantly between realizations. We repeated this test using the \axis\ geometry and changing the parameter $\sigma_{ring}$. The result is stable with respect to these assumptions. We found that the spatial scaling of the average E and B-mode power spectrum can be approximated by a decreasing power-law $\mathcal{D}_{\ell} \sim \ell^{\alpha}$, with $\alpha = -1.78$. \\*
Our simulations suggest that the level of polarized CO emission from unresolved clouds, despite being significantly lower than  synchrotron or thermal dust, can nevertheless significantly bias the primordial B-mode signal if not taken into account. The signal concentrates mainly on large angular scales and at $\ell\sim 80$, $\mathcal{D}_{\ell}=(1.1 \pm 0.8)\times 10^{-4} \mu \mathrm{K}^2$ where the uncertainty corresponds to the error in the mean spectra estimated from the 100 MC realizations. Therefore, the level of contamination is comparable to a primordial B-mode signal induced by tensor perturbations of amplitude $r_{CO}= 0.003 \pm 0.002$, i.e. below the recent upper limit $r<0.07$ reported by the BICEP2 Collaboration \citep{2016PhRvL.116c1302B} but higher than the $r=0.001$ target of upcoming experiments \citep{2016arXiv161002743A, 2014JLTP..176..733M, 2016arXiv161208270C}. The contamination quickly becomes  sub-dominant on small angular scales ($\ell\approx 1000$) where the B-modes are mostly sourced by the gravitational lensing.\\*

We finally note that these estimates are conservative since  the assumed polarization fraction of $1\%$ of polarized is close to the high end of the polarization fractions observed in CO clouds. 

\begin{figure}
\subcaptionbox{}[\columnwidth]{\includegraphics[width=.9\columnwidth]{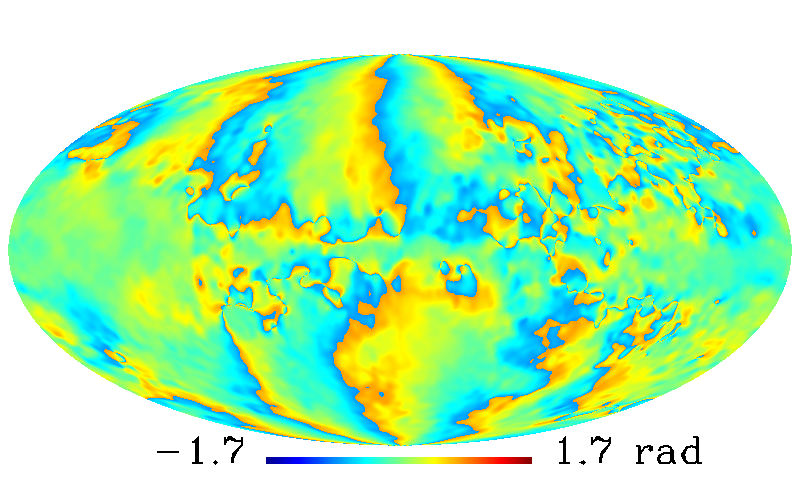}}\\
\subcaptionbox{}[\columnwidth]{\includegraphics[width=.9\columnwidth]{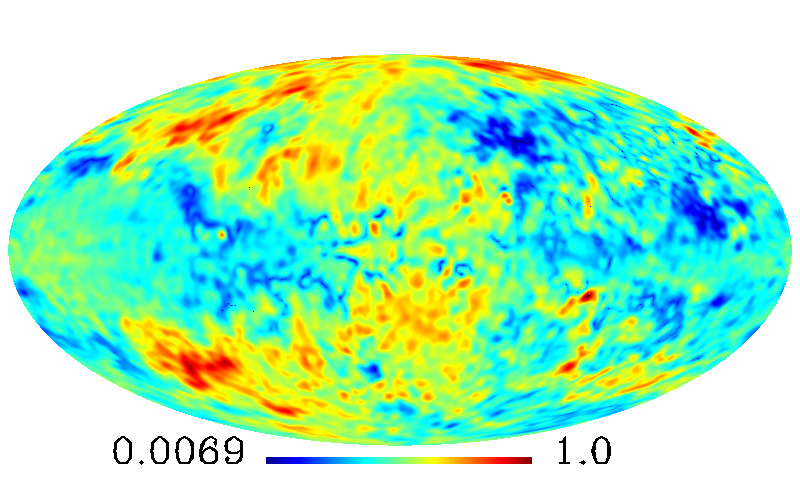}}
\caption{(a) Polarization angle and (b) geometrical depolarization maps used for simulating polarized CO emission in this work.}\label{fig:depolamap}
\end{figure}
\section{Conclusions}
In this work we have developed a parametric model for CO molecular line emission which takes account of the CO clouds distribution within our Galaxy in 3D with different geometries, as well as the most recent observational findings concerning their sizes, locations, and emissivity. \\*
Despite most of the observations having so far been confined to  the Galactic plane, we have built the model to simulate the emission over the full sky. The code implementing \mcmole\ is being made publicly available. \\*
We have compared the results of our simulations with \pl\ CO data on the map level and statistically (by matching angular power spectra). We found that:
\begin{enumerate}
\item the parameters of the size function, $L_0$, and the width of the Galactic radial distributions $\sigma_{ring}$ play a key role in shaping the power spectrum; 
\item the choice of symmetries in the cloud distribution changes the profile of the integrated emission in the Galactic plane (\autoref{eq:modela}) but not the power spectrum morphology;
\item our model is capable of reproducing fairly well the observations at low Galactic latitudes (see \autoref{fig:renorm}) and the power spectrum at high latitudes (\autoref{fig:bestfit_ht}).
\end{enumerate}
We used our model to fit the \pl\ observed CO power spectrum and to estimate the most relevant parameters of the CO distribution, such as the typical size of clouds and the thickness of the molecular ring, finding results in agreement with values reported in the  literature. 
The model which we have developed could easily be generalized and extended whenever new data become available. In particular, its accuracy at high Galactic latitudes would greatly benefit from better sub-mm measurements going beyond the Planck sensitivity, as well as from better information about the details of the CO polarization properties.\\*
Finally, we used the best-fit parameters obtained from comparing  the \mcmole\ model with \pl\ data to forecast the unresolved CO contamination of the \cmb\ B-mode power spectrum at high Galactic latitudes. We conservatively assumed a polarization fraction of $p_{CO}=1\%$, which corresponds to the high end of those observed at low latitudes,since no polarized CO cloud has yet  been observed far from the Galactic plane due to the weakness of this emission.\\*
We found that this signal could mimic a B-mode signal with tensor-to-scalar ratio $0.001\lesssim r\lesssim 0.025$.
This level of contamination is indeed relevant for accurate measurements of CMB B-modes. It should therefore be inspected further in light of the achievable sensitivities of upcoming and future CMB experiments together with the main diffuse polarized foreground (thermal dust and synchrotron). From the experimental point of view, trying to find  dedicated instrumental solution for minimizing the impact of CO emission lines, appears to be  particularly indicated in the light of these results. 

\section*{ }
{
\small
\emph{Acknowledgements.} We would like to thank Fran\c{c}oise Combe for many useful comments and suggestions for the development of this study, as well as  Alessandro Bressan, Luigi Danese, Andrea Lapi, Akito Kusaka, Davide Poletti and Luca Pagano for several enlightening discussions. We thank John Miller for his careful reading of this work. We thank Guillaume Hurier for several clarifications about the \pl\ CO products. 
This work was supported by the RADIOFOREGROUNDS grant of the European Union's Horizon 2020 research and innovation programme (COMPET-05-2015, grant agreement number 687312) and by  by Italian National Institute of Nuclear Physics (INFN) INDARK project. GF acknowledges support of the CNES postdoctoral programme. 
Some of the results in this paper have been derived using the HEALP{\scriptsize IX} \citep{2005ApJ...622..759G} package.
}

\bibliographystyle{plainnat}

\bibliography{co_bib}

\bsp	
\newpage
\label{lastpage}
\appendix

\section{Best-fit with \axis\ geometry}\label{app:bestfit}
\begin{figure*}
\includegraphics[width= .5\textwidth]{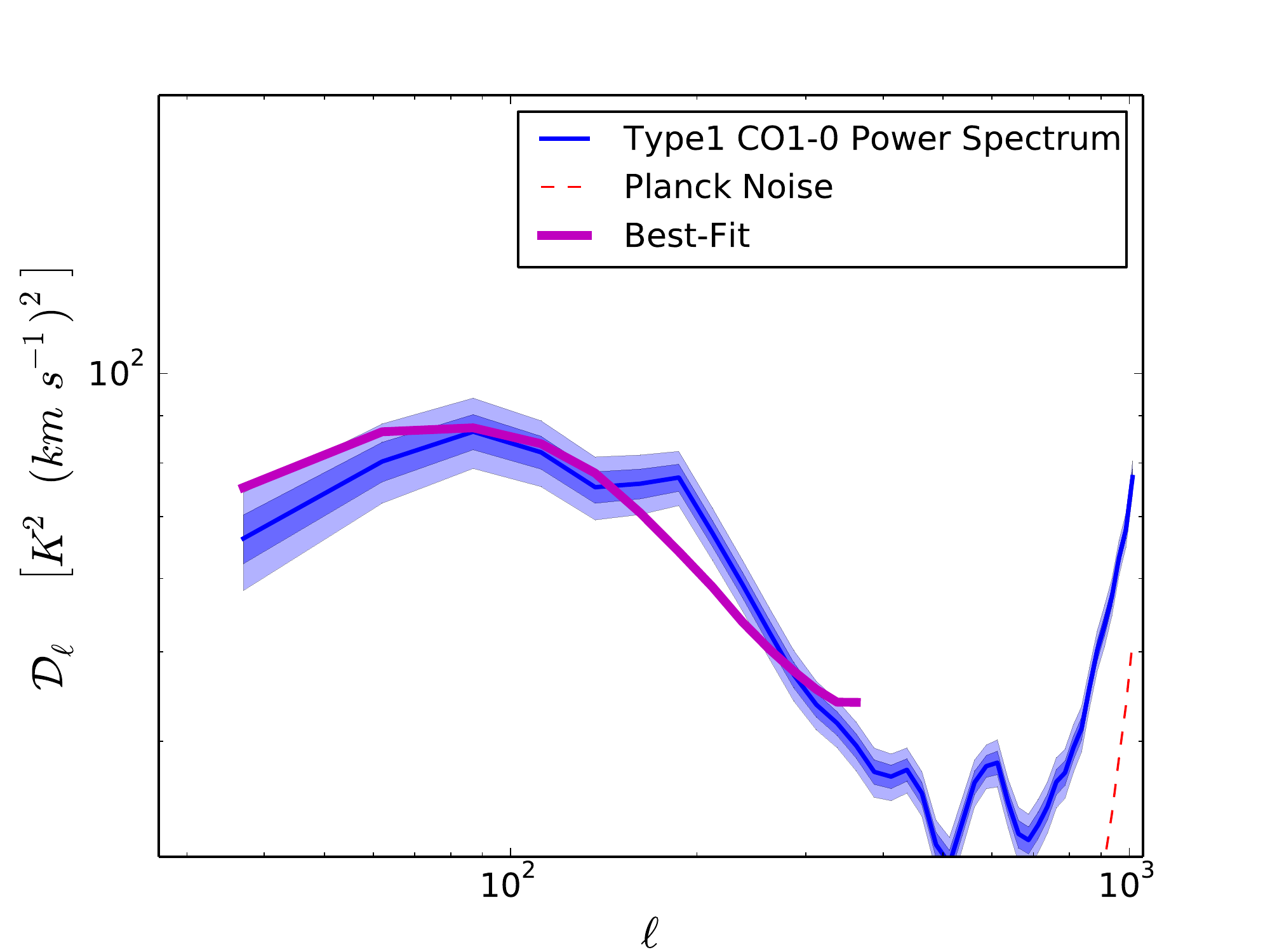}
\includegraphics[width=  .5\textwidth]{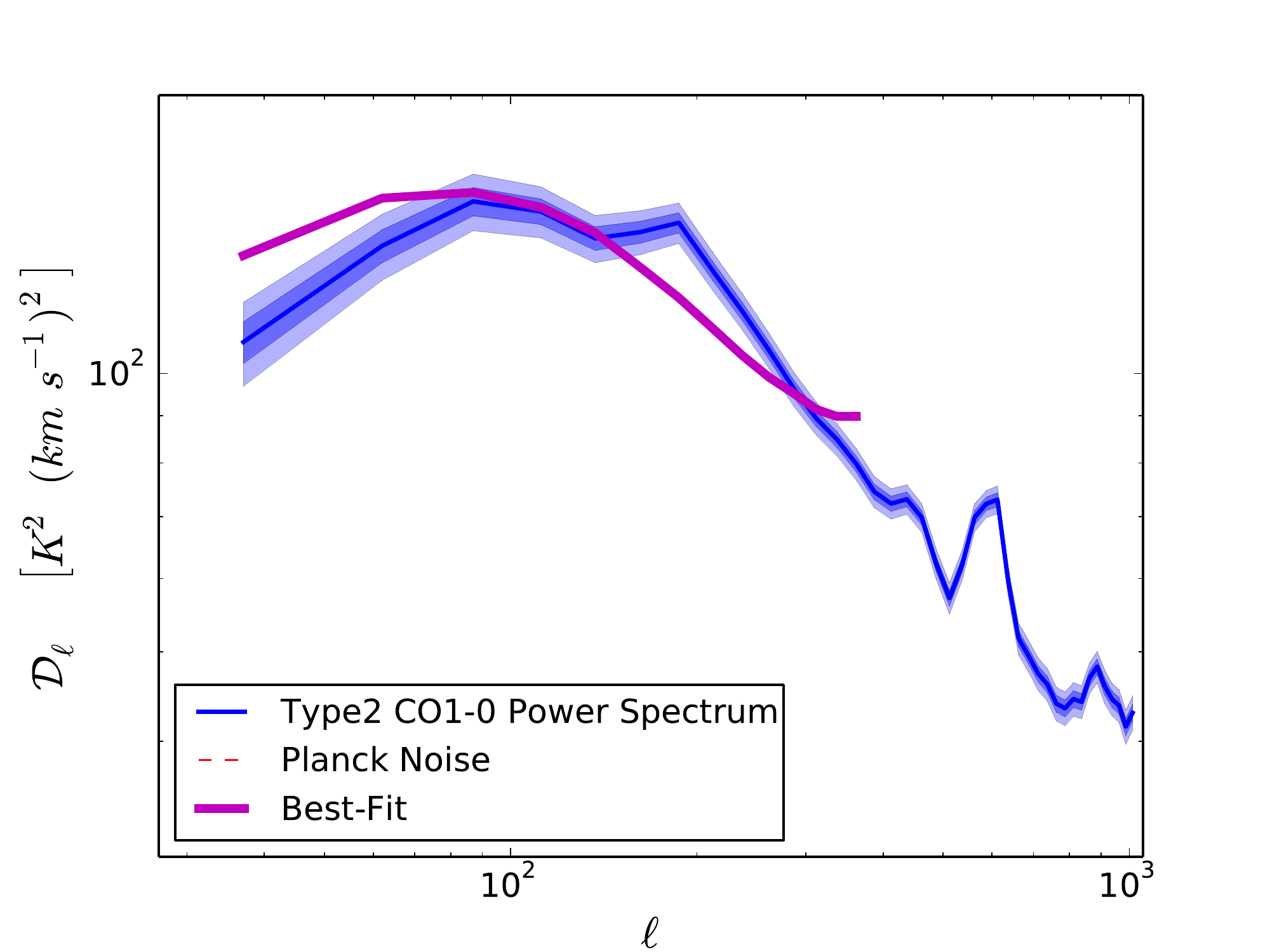}
\caption{ Angular power spectra of \pl\ CO $1-0$ line (blue) for \emph{Type 1} (left) and \emph{Type 2} (right) maps together with the results of the \mcmole\ best-fit model adopting an \axis\ geometry (magenta).}\label{fig:bestfit_axis}
\end{figure*}

\begin{figure*}
\includegraphics[width=   .5\textwidth]{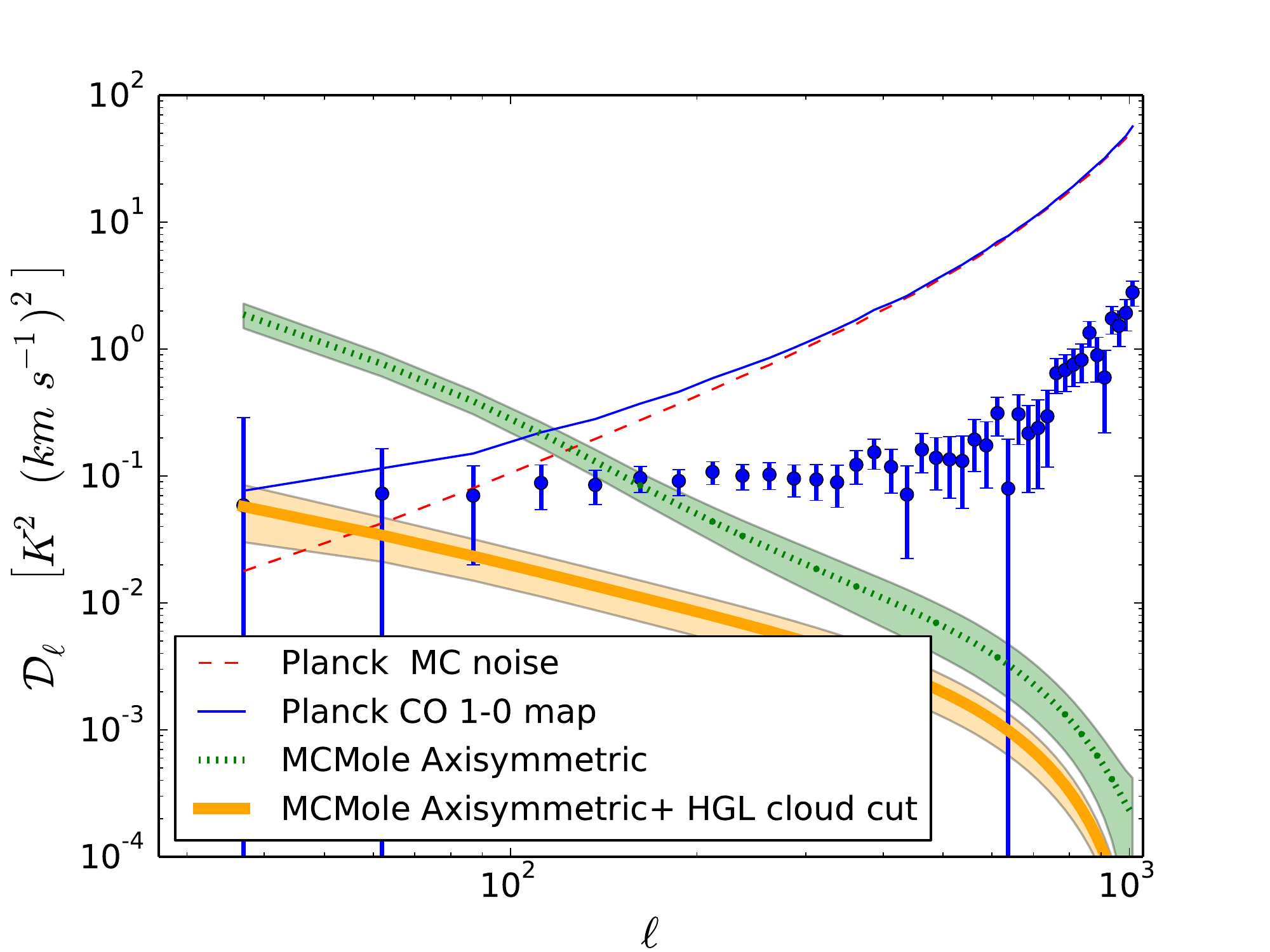}
\includegraphics[width=   .5\textwidth]{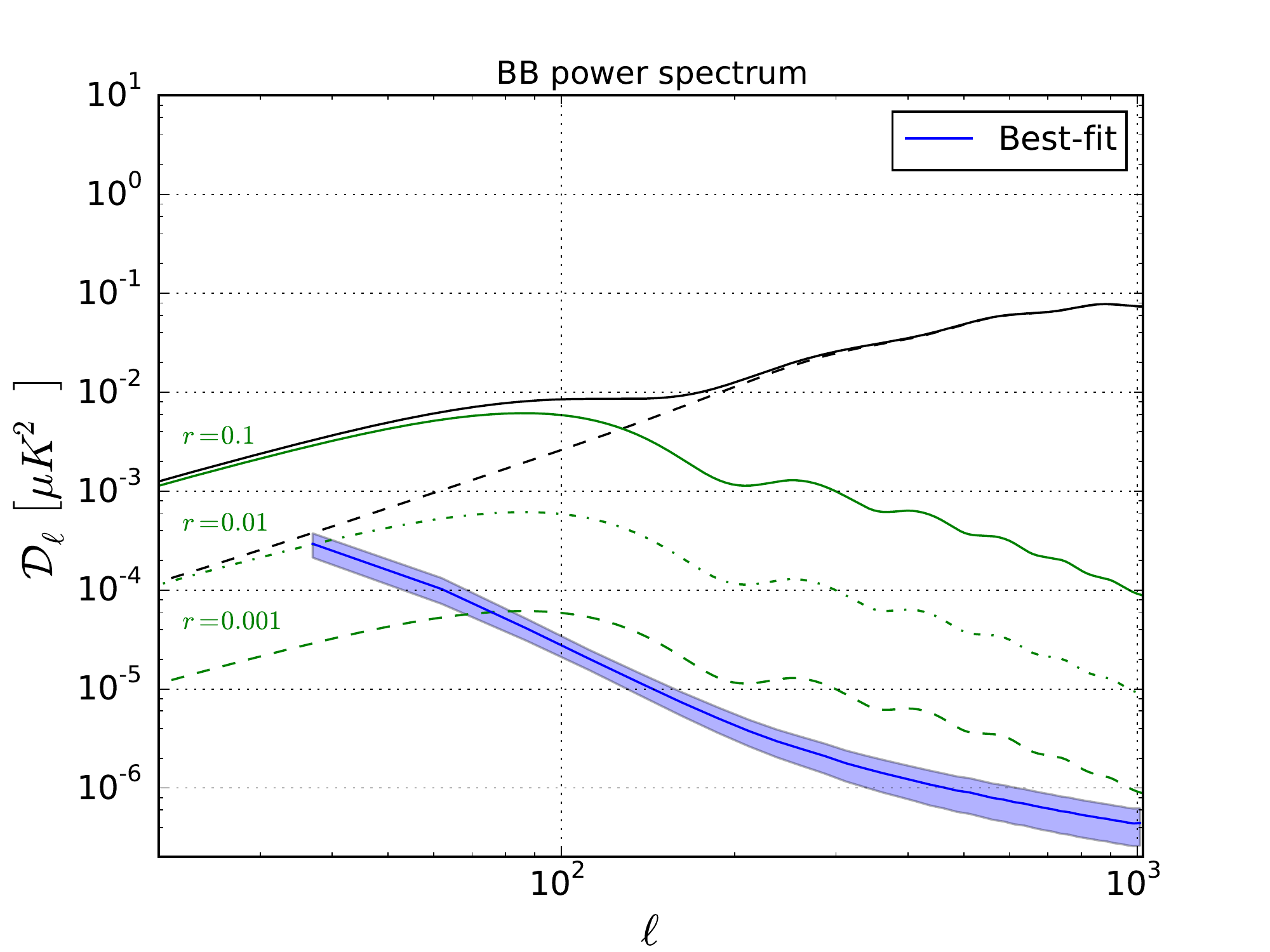}
\caption{ Left: CO 1-0 power spectrum at high Galactic latitudes of the \axis\ \mcmole\ model (dotted green), for the best-fit parameters reported in \autoref{tab:bestfit_axis}. Thick orange solid line shows the power spectrum for the  \axis\ geometry with the HGL-cut applied. The \pl\ \emph{Type 1} CO power spectrum before and after noise bias subtraction is shown  with the blue solid line and dots respectively. The \pl\ noise bias is shown with the dashed red line. Right:  B-mode power spectra of polarized CO emission lines at high Galactic latitudes estimated using the best-fit parameters of the \axis\  \mcmole\ model, see \autoref{fig:pol} for a comparison with the \spir\ geometry.}\label{fig:bestfit_ht_ax}
\end{figure*}

In this appendix we present the results of the analysis described in \autoref{sec:bestfit} to constraint the CO distribution using the \mcmole\ model adopting an \axis\ geometry instead of the \spir\ one. Following the procedure of \autoref{sec:bestfit} we construct a series of $\mathcal{F}(\ell; \sigma_{ring}, L_0)$ hyper-surfaces sampled on an ensemble of specific values of the $L_{0}$ and $\sigma_{ring}$  parameters within the same ranges reported in \autoref{sec:bestfit}.\\*
 In \autoref{fig:bestfit_axis} we show the results of the fit of the axisymmetric \mcmole\ model to the CO power spectrum of the \pl\ \emph{Type 1} and \emph{Type 2} CO maps in the Galactic plane. We summarize the best-fit values of these parameters in \autoref{tab:bestfit_axis}. As it can be seen from the results of the $\chi^2$ test in \autoref{tab:bestfit_axis} the \axis\ model does not fit the data satisfactorily. Moreover one of the parameters of the model, the typical cloud size $L_{0}$, is in practice unconstrained. For this reason we decided to adopt the \spir\ geometry as a baseline choice for our forecast presented in \autoref{sec:forecasts}. Nevertheless, we pushed the comparison between the two geometries in the high galactic latitude area for sake of completeness.\\* 
In \autoref{fig:bestfit_ht_ax} we show the comparison between the \pl\ data for \emph{Type 1} maps and the \mcmole\ axisymmetric best-fit model after the application of the HGL cut described in the paper. The \axis\ model describes the data similarly to the \spir\ model at the larger scales. The difference in the signal amplitude is in fact less then 30\% for angular scales $\ell\lesssim 100$ and the two models are compatible within the error bars. This seems to indicate that in this regime, the details of the CO distribution in the high galactic latitude region are mainly affected by the properties of the vertical profile rather than by the geometry of the distribution. Conversely, the difference between the two geometries becomes important at smaller angular scales reaching a level of $\approx 2$ at $\ell\approx 1000$. \\*
We finally performed a series of polarized simulations as in \autoref{subsec:pol-sim} to access the level of contamination to the CMB B-modes power spectrum with the best-fit \axis\ model and found $r_{CO}\lesssim0.001$.
 Moreover, the slope of the BB power spectrum in \autoref{fig:bestfit_ht_ax}(b) is $-2.2$ similar to the one computed with the \spir\ geometry.

 Because the \spir\ model describes the data both in the high and low galactic latitude area, we consider the upper limit derived with this setup more reliable and the reference estimate for the contamination to the cosmological signal due to the CO polarized emission.

\begin{table*}
\begin{tabular}{c c c c c c c c}
\toprule 
 & $L_0$ [pc]& $\sigma_{ring}$ [kpc] & $A_{CO}$& $\tilde{\chi}^2$ & \emph{dof} & $p$-value & $\rho_{L\sigma} $\\
\midrule
Type 1 & $19.47 \pm 12.68$ & $ 2.12 \pm 0.23$ & $1.00 \pm 0.12$ & $7.35$ &  11 & 0.00 &  0.99\\
Type 2 & $16.24 \pm 17.56$ & $2.12 \pm 0.30$  & $2.25 \pm 0.35$ & $18.08$ & 11 & 0.00&0.99 \\
\bottomrule
\end{tabular}
\caption{ Best-fit parameters for the \axis\ \mcmole\ model.}\label{tab:bestfit_axis}
\end{table*}

\end{document}